\documentclass[trackchanges]{aastex701}

\usepackage{amsmath}
\usepackage{booktabs}
\usepackage{rotating}

\begin{document}

\title{\texttt{pespace}: A new tool of GPU-accelerated and auto-differentiable response generation and likelihood evaluation for space-borne gravitational wave detectors}

\author[orcid=0000-0001-9098-6800,gname='Rui',sname='Niu']{Rui Niu}
\affiliation{Department of Astronomy, University of Science and Technology of China, Hefei 230026, China}
\affiliation{School of Astronomy and Space Sciences, University of Science and Technology of China, Hefei 230026, China}
\email{nrui@ustc.edu.cn}  


\author[orcid=0000-0001-7438-5896,gname='Chang',sname='Feng']{Chang Feng}
\affiliation{Department of Astronomy, University of Science and Technology of China, Hefei 230026, China}
\affiliation{School of Astronomy and Space Sciences, University of Science and Technology of China, Hefei 230026, China}
\email[show]{changfeng@ustc.edu.cn}  

\author[orcid=0000-0001-9098-6800,gname='Wen',sname='Zhao']{Wen Zhao}
\affiliation{Department of Astronomy, University of Science and Technology of China, Hefei 230026, China}
\affiliation{School of Astronomy and Space Sciences, University of Science and Technology of China, Hefei 230026, China}
\affiliation{College of Physics, Guizhou University, Guiyang 550025, China}
\email[show]{wzhao7@ustc.edu.cn}

\correspondingauthor{Chang Feng, Wen Zhao}

\begin{abstract}
Space-borne gravitational wave detectors will expand the scope of gravitational wave astronomy to the milli-Hertz band in the near future. The development of data analysis software infrastructure at the current stage is crucial to both forecasting studies of physical questions and investigation of new data analysis method prototypes.
{In this paper, we introduce \texttt{pespace} which can be used for the full Bayesian parameter estimation of massive black hole binaries with space-borne detectors including LISA, Taiji, and Tianqin.
The core computations are implemented using the high-performance parallel programming framework \texttt{taichi-lang}\footnote{The \href{https://www.taichi-lang.org/}{taichi} programming language \citep{Hu2019,Hu2019a} and the \href{https://taiji.ictp-ap.org/}{Taiji} mission \citep{Luo2020} share the same name in Chinese. To avoid confusion, we use the Wade–Giles romanization with typewriter font for \texttt{taichi-lang} to make a distinction.}
which enables automatic differentiation and hardware acceleration across different architectures.
We also reimplement the waveform models \texttt{PhenomXAS} and \texttt{PhenomXHM} in the separate package \texttt{tiwave} to integrate waveform generation within the \texttt{taichi-lang} scope, making the entire computation accelerated and differentiable.}
To demonstrate the functionality of the tool, we use a typical signal from a massive black hole binary to perform the full Bayesian parameter estimation with the complete likelihood function for three scenarios: including a single detector using the waveform with only the dominant mode; a single detector using the waveform including higher modes; and a detector network with higher modes included.
The results demonstrate that higher modes are essential in breaking degeneracies, and coincident observations by the detector network can significantly improve the measurement of source properties.
Additionally, automatic differentiation provides an accurate way to obtain the Fisher matrix without manual fine-tuning of the finite difference step size.
Using a subset of extrinsic parameters, we show that the approximated posteriors obtained by the Fisher matrix agree well with those derived from Bayesian parameter estimation.
We aim to provide a convenient and easy-to-use tool for the preparatory science of space-borne gravitational wave missions. More functionality will be continuously developed in future work.
\end{abstract}

\section{Introduction} \label{sec_intro}
\setcounter{footnote}{0}
The gravitational wave (GW) observations provide a new tool for investigating the nature of the Universe.
Currently, ground-based detectors have made remarkable achievements. More than 200 GW events have been detected in four scientific observing runs of the LIGO-Virgo-KAGRA Collaboration (LVK) to date, which have substantially enriched our understanding of astrophysics, cosmology, and theories of gravity \citep{LVKCollaboration2019,LVKCollaboration2020b,LVKCollaboration2021e,LVKCollaboration2021d,LSC2025}.
In the near future, space-borne detectors including the proposed LISA \citep{AmaroSeoane2017a}, Taiji \citep{Luo2020}, and Tianqin \citep{Luo2016} missions will further open the window of GWs in the milli-Hertz band where a wide variety of GW sources exist, including massive black hole binaries (MBHBs), galactic compact binaries, stellar origin black hole binaries, extreme mass ratio inspirals, stochastic backgrounds, and potential unmodeled sources \citep{AmaroSeoane2023,Ruan2020}. Space-borne detectors will extend the scope of GW astronomy to the low-frequency regime beyond the reach of ground-based detectors and provide complementary and new views of the Universe \citep{Baker2019}.

The diverse and numerous signals in the milli-Hertz band also pose new challenges for the data analysis of space-borne detectors \citep{Speri2022a}.
Due to the correlations among the overlapping signals, global-fitting \citep{Cornish2007,Littenberg2020,Littenberg2023,Karnesis2023,Katz2024} where all parameters are inferred simultaneously has to be used for parameter estimation, which is computationally expensive and complicated to implement. 
Additionally, the Whittle likelihood \citep{LVKCollaboration2020f} under the stationary Gaussian noise assumption commonly used currently for ground-based detectors may not be adequate for handling data from space-borne detectors with nonstationary noise \citep{Cornish2020,Digman2022a,Du2025}, data gaps \citep{Baghi2019,Mao2024,Castelli2024,Wang2024b,Burke2025}, glitches \citep{Spadaro2023a,Baghi2022a}, etc.
These issues are still under active investigation currently, and an easy-to-use tool for basic functions like generating detector responses can facilitate studies on unsolved questions.
On the other hand, for the forecasting research on physical questions, a rough estimation of parameter measurement uncertainties is usually sufficient to assess the ability of future observations to constrain theoretical models \citep{Klein2016,Niu2020,Gao2023a,Lyu2025,Che2025,Baker2022a}.
Thus, for the preparatory science of future space-borne detectors, it is still reasonable to develop the tool for parameter estimation with the ideal stationary Gaussian noise at the current stage.
Although the global-fitting is required for overlapping signals of space-borne detectors, the correlations among different types of GW sources are modest. The Gibbs update scheme \citep{Casella1992,Gelfand2000} is a practical approach to address the extreme high-dimensional parameter space of global-fitting, where random sampling is performed sequentially for different source types \citep{Littenberg2020,Katz2024}.
Therefore, parameter estimation tools for different source types can be developed separately, and assembled by the Gibbs scheme to obtain the joint posterior distribution.
In this work, we focus primarily on signals from MBHB systems, and other source types and implementation of the global-fitting will be developed in future work.

Contemporary high-performance computing hardware offers new possibilities for addressing challenges in GW data analysis and makes previously computationally prohibitive approaches feasible in practice \citep{Santini2025,GarciaQuiros2025,Bandopadhyay2024b,Speri2023,Saltas2023,Strub2023,Katz2020,Katz2022a,Wysocki2019}.
Exploiting the latest developments in computing hardware and software can benefit the design and implementation of high-performance data analysis pipelines for next-generation GW observation missions.
However, the GPU acceleration through low-level hardware-oriented programming frameworks like \texttt{CUDA} is usually complex to implement, and extensive profiling and optimizing efforts are required for substantial performance gains. 
Modern numerical computation frameworks such as \texttt{jax}, \texttt{warp}, and \texttt{taichi-lang} provide promising alternative approaches to leverage contemporary high-performance computing hardware.
These frameworks provide abstractions for high-level programming, manage low-level optimizations internally, and allow users to focus on physical modeling rather than hardware details \citep{Frostig2018,Hu2019}.
In addition, automatic differentiation \citep{Corliss2006,Baydin2015,Margossian2019} has become an increasingly important tool in data analysis workflows. It enables efficient and accurate evaluation of gradients or partial derivatives, which are essential for gradient-based sampling or optimization algorithms \citep{Betancourt2017,Hoffman2011,Nocedal2006}, as well as for the Fisher forecasting analysis \citep{Iacovelli2022,Iacovelli2022b}.
The advantages of hardware acceleration and automatic differentiation have been demonstrated in data analysis for ground-based detectors \citep{Edwards2024, Wong2022a, Iacovelli2022, Iacovelli2022b}, where the differentiable waveform \texttt{ripple}, the novel gradient-based sampler \texttt{flowMC}, and the Fisher matrix package for GW cosmology \texttt{gwfast} have been developed.

The prohibitively high computational cost is one of the obstacles to perform full Bayesian parameter estimation.
This issue is particularly pronounced for space-borne detectors, owing to more complex response models and longer duration signals.
Two different strategies can be adopted to address this issue in parameter estimation for MBHB signals with space-borne detectors.
One strategy is developing more efficient stochastic sampling algorithms that can reduce the number of likelihood evaluation required to obtain posteriors, or circumventing the computationally expensive likelihood evaluation by using simulation-based inference methods or by constructing low computational cost surrogates for likelihood or posterior \citep{Jan2024,Gammal2025,Sharma2024,Hoy2024,Vilchez2024,Du2023,Ruan2023}.

The other strategy focuses on accelerating individual likelihood evaluation by simplifying the likelihood function under approximations or by leveraging contemporary computational hardware to reduce the wall-time for likelihood computation, which is also adopted in this work.
Currently, there are several existing tools that offer similar functionalities for generating detector responses and evaluating likelihood using this approach, including
\begin{itemize}
    \item \href{https://github.com/eXtremeGravityInstitute/LISA-Black-Hole}{\texttt{LISA-Black-Hole}}\citep{Cornish2020a} uses a reference signal close to the true signal to simplify the likelihood by the technique of heterodyning \citep{Cornish2021c}. By using the residual after subtracting the reference signal, the likelihood can be separated into three parts including a constant that is independent of the signal parameters and can be computed before parameter estimation, a part slowly varying which can be computed using interpolation on a coarse frequency samples, and a part rapidly oscillating but with an exponentially damped envelope which can be computed by a small fraction of the full extent. These measures make the likelihood inexpensive to evaluate and reduce the cost of full Bayesian parameter estimation.
    
    \item \href{https://gitlab.in2p3.fr/marsat/lisabeta}{\texttt{lisabeta}}\citep{Marsat2021} adopts the noise-free likelihood where the noise realization is not considered. By using the well-designed interpolation strategy in which the amplitude and phase of the signal are represented on a sparse frequency grid with different orders of splines, the computational cost of likelihood evaluation can be largely reduced.
    
    \item \href{https://github.com/mikekatz04/BBHx}{\texttt{bbhx}} \citep{Katz2020} also provides similar functionality for accelerating likelihood evaluation. The core computations in \texttt{bbhx} are implemented with \texttt{CUDA}, which enables the GPU acceleration and makes the full Bayesian parameter estimation using the complete likelihood without any simplification to be computationally feasible in practice.
\end{itemize}
Based on these existing tools, extensive studies performing full Bayesian parameter estimation for signals from MBHB systems have been conducted to investigate the development of data analysis pipeline prototypes for future observations \citep{Wang2022,Katz2022,Cornish2022,Gao2025,Weaving2023,Hoy2023,Chen2023,Li2023,Du2025a} and forecast the capabilities of future observations for addressing scientific questions such as testing gravitational theories \citep{Niu2024,Piarulli2025}, and measuring astrophysical properties of MBHBs \citep{Garg2024b,Garg2023a}.

{In this work, we provide new tools \href{https://github.com/nnrui/tiwave}{\texttt{tiwave}} and \href{https://github.com/nnrui/pespace}{\texttt{pespace}} to generate waveforms, compute detector responses, and evaluate likelihood for space-borne detectors. 
Compared with previous tools above, the main improvements include the following: the detector network of LISA, Taiji, and Tianqin is incorporated and can be used jointly for parameter estimation; the waveform models of \texttt{PhenomXAS} \citep{Pratten2020} and \texttt{PhenomXHM} \citep{GarciaQuiros2020} which are considered to exhibit better agreement with numerical relativity waveforms than \texttt{PhenomD} and \texttt{PhenomHM} used in previous tools are supported\footnote{Currently, only the configuration of default \href{https://git.ligo.org/lscsoft/lalsuite/-/blob/e59d1f0003d2358dd49b9795073624fdcf1ce0a8/lalsimulation/lib/LALSimIMRPhenomX.c\#L136}{waveform flags} is implemented, except for the setting of multibanding which is not supported at present. In all performance tests and validation tests discussed in Sec.~\ref{subsec_perf}, App.~\ref{app_mismatch_32}, and \ref{app_mod_xas}, waveforms generated by \texttt{lalsimulation} use the default settings with multibanding disabled.}, and the core computations are implemented with \texttt{taichi-lang} which enables automatic differentiation and hardware acceleration across various architectures.
}
To demonstrate the basic functions of \texttt{pespace}, we perform the full Bayesian parameter estimation with a GW signal from a typical MBHB system, and demonstrate the significant improvement in parameter measurements achieved by incorporating the network of multiple detectors and higher modes of the waveform. 
Automatic differentiation allows efficient and accurate computation of partial derivatives for detector responses using the forward mode, which can facilitate the computation of the Fisher matrix.
The Fisher method provides a good approximation of the posterior by a multivariate Gaussian distribution under the large signal-to-noise ratio (SNR) assumption without requiring computationally expensive stochastic sampling \citep{Cutler1994}, which is particularly useful in forecast studies or initializing proposal distributions to improve the efficiency of stochastic sampling.
However, the Fisher matrix obtained by the finite difference method requires manually tuning difference steps to find the trade-off between truncation errors and round-off errors. This issue can be avoided by the automatic differentiation system of \texttt{taichi-lang} which can provide derivative computations with accuracy close to machine precision.
In this work, we use a subset of extrinsic parameters to demonstrate the agreement between approximated posteriors sampled from the multivariate Gaussian distribution given by the Fisher method and posteriors obtained from stochastic sampling using the noise-free likelihood.

It is also important to emphasize that our work has several known limitations, which also outline the roadmap for future developments.
First, we only consider individual MBHB signals here and focus on parameter estimation. Searches and the incorporation of other source types into the global-fitting framework are left to future work.
Second, the analysis in this work adopts idealized noise models where issues that may be encountered in real observed data such as non-stationary noise, glitches, data gaps are not taken into account.
Third, we employ the frequency domain response model based on the stationary phase approximation, which is applicable to signals chirping fast enough like MBHBs. Other source types may not be directly compatible with this formalism for constructing frequency domain responses, and applicability to effects like double-spin precession \citep{Chatziioannou2017} still requires further investigation.
Finally, the forward-mode automatic differentiation which is required in situations involving large-dimensional outputs and small-dimensional inputs currently is not supported for waveform generation. Thus partial derivatives of detector responses with respect to intrinsic parameters of GW signals still need to be obtained via numerical differentiation. 

The remainder of this paper is organized as follows. In Sec.~\ref{sec_method}, we provide a brief review of the frequency domain response model used in this work and the Bayesian framework for parameter estimation. Implementation details are also presented in this section.
Using a signal of a typical MBHB system, we perform the full Bayesian parameter estimation using the complete likelihood without simplification for three scenarios. We also consider a subset of extrinsic parameters as the example to demonstrate the agreement between posteriors approximated by the Fisher method and posteriors obtained from stochastic sampling. The obtained results are shown in Sec.~\ref{sec_results}.
Finally, Sec.~\ref{sec_summary} provides a summary of this work and discusses perspectives on future work.
The code and data to reproduce all results shown in this paper are available at \url{https://doi.org/10.5281/zenodo.18339164}. Two core packages used in this work \texttt{pespace} and \texttt{tiwave} are available at \url{https://github.com/nnrui/pespace} and \url{https://github.com/nnrui/tiwave}.

\section{Methodology} \label{sec_method}

\subsection{Response model}\label{subsec_response}
Space-borne detectors exhibit many differences in their responses to GWs \citep{Cornish2003,Vallisneri2005,Rubbo2004,Krolak2004,Hu2018} compared with currently operating ground-based detectors owing to longer arm-length, orbital motion, unequal arm-length, etc.
For ground-based detectors, the characteristic size is much less than the length-scale where the gravitational field changes substantially for GWs in the sensitive band. Thus, one can work under the long wavelength approximation with the proper detector frame \citep{Maggiore2007a}. The detector responses to GWs can be obtained by assembling the GW waveform with the pattern functions which are constants depending on the source direction, the detector geometry, and the GW polarization.
However, for space-borne detectors the long wavelength approximation is not strictly valid. For example, the proposed design of LISA has the arm-length of $L=2.5\times 10^6\mathrm{km}$ corresponding to the transfer frequency $f_*\equiv1/(2\pi L)\sim0.019\mathrm{Hz}$ which is still within the sensitive range of space-borne detectors.
During the propagation of laser photons from emission to reception, there may be GWs of multiple cycles passing through the path of photons.
The crests and troughs of GWs can cancel mutually, which deteriorates the responses of detectors to GWs.
Therefore, the detector response becomes frequency-dependent, rather than being constant and determined solely by geometric angles.

Another important difference of space-borne detectors compared with ground-based detectors is that the orbital motion has to be accounted for.
GW signals can linger within the sensitive band of space-borne detectors for months or years \citep{AmaroSeoane2023}. The orbital motion of detectors can induce time-dependent modulations and delays in the response, which complicate the Fourier transform to obtain an analytic form of response in the frequency domain.

Additionally, space-borne detectors are unequal-arm interferometers where the laser frequency fluctuations will experience different delays when propagating along two arms. 
Thus the laser frequency noise which can be several orders of magnitude stronger than GW signals cannot be mutually canceled by taking the difference at the photodetector. 
The post-processing technique of time delay interferometry (TDI) which can suppress the laser frequency noise by applying time-shifts and linear combinations on independent readouts of single laser links has to be used to construct the actual measurements \citep{Tinto2020}.

Below, we provide the detailed formulas for the detector responses in the Fourier domain following \citep{Marsat2018,Marsat2021}.
Working under the approximations including: 
the constellation of detector is a rigid equilateral triangle with an equal and constant arm-length $L$;
the Doppler effect induced by the relative speeds of spacecraft can be neglected; 
the motion of spacecraft during the laser propagation is small enough thus the point-ahead effect is ignored; 
the curvatures of spacetime induced by the Sun and other celestial bodies in the solar system are negligible, 
and using the unit of the fractional laser frequency shift, the response of a single laser link from the node $s$ to the node $r$ can be given by
\begin{equation} \label{eq_td_resp}
    y_{rs}(t) \equiv \frac{\nu_r - \nu_s}{\nu_s} = \frac{1}{2(1-\boldsymbol{k}\cdot\boldsymbol{n}_{rs})}
    \boldsymbol{n}_{rs}\cdot
    \bigg[
    \boldsymbol{h} (t-L-\boldsymbol{k}\cdot\boldsymbol{x}_s) - \boldsymbol{h} (t-\boldsymbol{k}\cdot\boldsymbol{x}_r)
    \bigg]
    \cdot\boldsymbol{n}_{rs},
\end{equation}
where $h(t)$ is the GW waveform, unit vectors $\boldsymbol{k}$ and $\boldsymbol{n}_{rs}$ denote directions of the GW propagation and the laser link, $\boldsymbol{x}_s$ and $\boldsymbol{x}_r$ represent positions of spacecraft for the sending node and the receiving node.
Here, we use idealized analytic orbital models for these vectors.
The Keplerian geocentric orbit \citep{Hu2018} is adopted for Tianqin, and the Keplerian heliocentric orbits \citep{Stas2020,Ren2023} with different initial values are used for LISA and Taiji.

The TDI observables are constructed by time shifting and linear combining of single link responses. 
Various forms of the TDI combination are still under active investigation \citep{Wang2025,Tan2025,Wang2024d,Wang2024c,Wang2025a}, and diverse new techniques have been proposed, such as $\text{TDI-}\infty$ \citep{Houba2024} and Bayesian TDI \citep{Page2021}.
In this work, we only consider the most commonly used 1.5- and 2.0-generation Michelson combinations \citep{Babak2021}.
The first generation combination is constructed under the assumption that the constellation is rigid and static, which is rarely used in recent studies. The 1.5-generation combination keeps the assumption of rigid constellation but consider the rotation which leads delays of different directions for the same link to be non-commutative, i.e., $L_{rs}=\mathrm{const}$, but $L_{rs}\neq L_{sr}$. 
The 2.0-generation combination further accounts for the flexing motion of spacecraft by considering a time-dependent arm-length with the linear order correction, i.e., $L_{rs}(t)=L_{rs}+\dot{L}_{rs}t$. It has been shown that the linear correction can already adequately suppress the laser noise below the secondary noise of current designs, and the correction of acceleration can be neglected \citep{Tinto2020}.

Since detector readouts have to be processed using the TDI combination for obtaining observables, we also need to apply the same TDI combination to single link responses to get responses of corresponding TDI channels.
However, the rotation and flexing of the constellation are only considered in derivation of TDI combinations for suppressing laser frequency noise, the rigid static equilateral triangle approximation is still adequate to be used for getting detector responses to GW signals. 
For the detector design including three spacecraft and six links, three TDI observables $X$, $Y$, $Z$ can be constructed.
With the delay operator defined as $\mathcal{D}_{rs}y(t)\equiv y(t-L_{rs})$, the 1.5- and 2.0-generation TDI combinations are given by \citep{Babak2021}
\begin{equation} \label{eq_td_TDI}
\begin{aligned}
    X_{1.5} 
    &= y_{13} + \mathcal{D}_{13}y_{31} + \mathcal{D}_{13}\mathcal{D}_{31}y_{12} + \mathcal{D}_{13}\mathcal{D}_{31}\mathcal{D}_{12}y_{21} \\
    &\phantom{=}
    -y_{12} - \mathcal{D}_{12}y_{21} - \mathcal{D}_{12}\mathcal{D}_{21}y_{13} - \mathcal{D}_{12}\mathcal{D}_{21}\mathcal{D}_{13}y_{31}, \\
    X_{2.0} 
    &= y_{13} + \mathcal{D}_{13}y_{31} + \mathcal{D}_{13}\mathcal{D}_{31}y_{12} + \mathcal{D}_{13}\mathcal{D}_{31}\mathcal{D}_{12}y_{21} \\
    &\phantom{=}
     + \mathcal{D}_{13}\mathcal{D}_{31}\mathcal{D}_{12}\mathcal{D}_{21}y_{12} + \mathcal{D}_{13}\mathcal{D}_{31}\mathcal{D}_{12}\mathcal{D}_{21}\mathcal{D}_{12}y_{21}\\
    &\phantom{=}
     + \mathcal{D}_{13}\mathcal{D}_{31}\mathcal{D}_{12}\mathcal{D}_{21}\mathcal{D}_{12}\mathcal{D}_{21}y_{13}\\
    &\phantom{=}
     + \mathcal{D}_{13}\mathcal{D}_{31}\mathcal{D}_{12}\mathcal{D}_{21}\mathcal{D}_{12}\mathcal{D}_{21}\mathcal{D}_{13}y_{31}\\
    &\phantom{=}
    - y_{12} - \mathcal{D}_{12}y_{21} - \mathcal{D}_{12}\mathcal{D}_{21}y_{13} - \mathcal{D}_{12}\mathcal{D}_{21}\mathcal{D}_{13}y_{31} \\
    &\phantom{=}
     - \mathcal{D}_{12}\mathcal{D}_{21}\mathcal{D}_{13}\mathcal{D}_{31}y_{13} - \mathcal{D}_{12}\mathcal{D}_{21}\mathcal{D}_{13}\mathcal{D}_{31}\mathcal{D}_{13}y_{31} \\
     &\phantom{=}
     - \mathcal{D}_{12}\mathcal{D}_{21}\mathcal{D}_{13}\mathcal{D}_{31}\mathcal{D}_{13}\mathcal{D}_{31}y_{12} \\
   &\phantom{=}
     - \mathcal{D}_{12}\mathcal{D}_{21}\mathcal{D}_{13}\mathcal{D}_{31}\mathcal{D}_{13}\mathcal{D}_{31}\mathcal{D}_{12}y_{21}.
\end{aligned}
\end{equation}
The other two channels, $Y$ and $Z$, can be obtained by cyclic permutation of indices.
Assuming that the noise in each link has identical properties and is uncorrelated, the orthogonal combinations can be constructed by 
\begin{equation}
    \begin{aligned}
        A &= \frac{1}{\sqrt{2}}(Z - X), \\
        E &= \frac{1}{\sqrt{6}}(X - 2Y + Z), \\
        T &= \frac{1}{\sqrt{3}}(X + Y + Z). \\
    \end{aligned}
\end{equation}

As will be discussed in the next subsection, for Gaussian and stationary noise, the covariance matrix of data in the Fourier domain is diagonal, and its inverse can be easily obtained in likelihood evaluation. Thus, the Bayesian parameter estimation is usually performed in the frequency domain. The analytic form of detector responses in the frequency domain is desirable to avoid the computational cost of the Fourier transforms.
For GW signals from MBHBs that chirp fast enough, as
shown in \citep{Marsat2018}, at leading order approximation, the time delay terms in Eq.~\ref{eq_td_resp} can be treated as a constant and the time-shift property of Fourier transform, $y(t-t_0)\xrightarrow{\mathrm{FT}}e^{-i2\pi f t_0}\tilde{y}(f)$, can be applied.
The time dependence in vectors associated with the geometry of constellation can be substituted by the time-to-frequency correspondence in the stationary phase approximation given by
\begin{equation}\label{eq_tf}
    t_f = -\frac{1}{2\pi} \frac{\mathrm{d} \Phi(f)}{\mathrm{d} f},
\end{equation}
where $\Phi(f)$ is the phase of the GW waveform.
The single link response in the frequency domain takes the form of
\begin{equation} \label{eq_fd_resp}
    \tilde{y}_{rs}(f) =
    -i\pi f L \ 
    \mathrm{sinc} \bigg[\pi f L (1-\boldsymbol{k}\cdot\tilde{\boldsymbol{n}}_{rs})\bigg]
    \exp 
    \bigg\{ -i \pi f 
        \bigg[L + \boldsymbol{k}\cdot(\tilde{\boldsymbol{x}}_r + \tilde{\boldsymbol{x}}_s) 
        \bigg] 
    \bigg\}
    \tilde{\boldsymbol{n}}_{rs}\cdot\tilde{\boldsymbol{h}} \cdot\tilde{\boldsymbol{n}}_{rs},
\end{equation}
where $\tilde{\boldsymbol{n}}_{rs}$, $\tilde{\boldsymbol{x}}_r$, and $\tilde{\boldsymbol{x}}_s$ are vectors of detector geometry in frequency domain which are obtained by replacing $t$ with $t_f$, $\tilde{\boldsymbol{h}}$ is the GW waveform in frequency domain.
Here, we use the usual convention for the Fourier transform, thus there is a difference of complex conjugation from the reference \citep{Marsat2018}.
In the frequency domain, the delay operator can be treated as the delay factor defined by $z=e^{-i2\pi f L}$ under the rigid static approximation. This allows the frequency domain TDI combinations to be reduced to
\begin{equation}
\begin{aligned}
    \tilde{X}_{1.5} &= (1-z^2)\big[\tilde{y}_{13} - \tilde{y}_{12} + z(\tilde{y}_{31} - \tilde{y}_{21})\big], \\
    \tilde{X}_{2.0} &= (1-z^2 -z^4 + z^6 )\big[\tilde{y}_{13} - \tilde{y}_{12} + z(\tilde{y}_{31} - \tilde{y}_{21})\big].
\end{aligned}
\end{equation}

\subsection{Parameter estimation} \label{subsec_pe}
With the response model $M$ introduced in the last subsection, for the given observed data $\boldsymbol{d}$, the probability distribution of parameters $\boldsymbol{\theta}$ describing the model $M$ can be given by Bayes' theorem \citep{Thrane2019,LVKCollaboration2020f},
\begin{equation}
p(\boldsymbol{\theta}|\boldsymbol{d}, M) \propto \pi(\boldsymbol{\theta}|M) p(\boldsymbol{d}|\boldsymbol{\theta}, M),
\end{equation}
where $\pi(\boldsymbol{\theta}|M)$ is the prior representing our knowledge of properties of the GW source before the observation, and $p(\boldsymbol{d}|\boldsymbol{\theta}, M)$ is the likelihood function representing the probability of noise fluctuations that happen to produce the observed data in the presence of a signal described by $\boldsymbol{\theta}$.

The explicit form of the likelihood function depends on the noise model. Here, we consider ideal stationary Gaussian noise, thus the observed data obey the multivariate Gaussian distribution
\begin{equation}
    p(\boldsymbol{d}|\boldsymbol{\theta}, M) \propto \exp \left[ 
    \sum_{i, j} -\frac{1}{2}\bigg(d(t_i)-s(t_i, \boldsymbol{\theta})\bigg)\Sigma^{-1}_{ij}\bigg(d(t_j)-s(t_j,\boldsymbol{\theta})\bigg)
    \right],
\end{equation}
where $s(t,\boldsymbol{\theta})$ denotes the responses of GW signal with parameters $\boldsymbol{\theta}$ for the TDI channel, subscripts $(i,j)$ are indices of time samples, and $\Sigma_{ij}$ is the covariance matrix of noise.
The size of the covariance matrix is determined by the number of time samples, and its inverse is impractical to compute for long duration observations.
However, if the noise is stationary, i.e., the correlations of noise between two time samples depend solely on their interval and are independent of their specific time instants, the covariance matrix takes a Toeplitz form and can be approximately diagonalized by the Fourier transform,
\begin{equation} \label{eq_diag_Sigma_to_Sn}
    \tilde{\Sigma}_{ij} = \frac{1}{2}S_n(f_i) T\delta_{ij},
\end{equation}
where $T$ is the duration of data, $S_n(f)$ is the power spectral density of noise.
In practice, the parameter estimation is usually performed in the frequency domain with the likelihood taking the form of
\begin{equation} \label{eq_ll_fd}
p(\boldsymbol{d}|\boldsymbol{\theta}, M) \propto \exp \left[
\sum_i -\frac{2 |\tilde{d}(f_i) - \tilde{s}(f_i,\boldsymbol{\theta})|^2}{S_n(f_i) T}
\right].
\end{equation}
For multiple independent observations $\boldsymbol{d}_{\{I\}}$, such as data from orthogonal TDI channels or different detectors, the joint likelihood is the product of individual likelihoods
\begin{equation}
p(\boldsymbol{d}_{\{I\}}|\boldsymbol{\theta}, M) = \prod_{I} p(\boldsymbol{d}_I|\boldsymbol{\theta}, M).
\end{equation}

Obtaining the posterior distribution in full Bayesian parameter estimation requires sufficient exploration of the entire parameter space and extensive likelihood evaluation which are usually computationally expensive.
However, as demonstrated in \citep{Vallisneri2008}, the likelihood can be expanded as a series in $1/\rho$ where $\rho$ denotes the optimal SNR of the fiducial signal or the best-fit signal.
In the high-SNR regime, where $1/\rho\to0$, by keeping only the leading order, the likelihood can be approximated as
\begin{equation} \label{eq_ll_fisher}
    p(\boldsymbol{d}|\boldsymbol{\theta}, M) \propto 
    \exp \left[\sum_{i,j}-\frac{1}{2} \delta \theta_i \delta \theta_j  F_{ij}\right],
\end{equation}
where $\delta \theta_i = \theta_i-\theta_{0,i} $ with $\theta_{0,i}$ denoting parameters of the fiducial signal, and $F_{ij}$ is the Fisher matrix given by
\begin{equation}
    F_{ij} = \left\langle \frac{\partial \tilde{s}}{\partial \theta_i} \middle| \frac{\partial \tilde{s}}{\partial \theta_j}\right\rangle.
\end{equation}
The notation $\langle\cdot|\cdot\rangle$ is defined by the noise weighted inner product as
\begin{equation} \label{eq_inner_product}
    \langle a | b \rangle = 4 \mathrm{Re} \sum_i \frac{a^*(f_i)b(f_i)}{S_n(f_i)T}.
\end{equation}
Considering uniform priors, the posterior distribution can be approximated by a multivariate Gaussian distribution with means $\theta_{0,i}$ and the covariance matrix $C_{ij} = F^{-1}_{ij}$ using the likelihood of Eq.~\ref{eq_ll_fisher}. This approximated posterior can be used for forecasting the measurement uncertainty of source parameters or constructing proposal distributions from results of a coarse parameter search to improve the efficiency of fine Bayesian parameter estimation. 
Recent improvements to the Fisher method, such as the derivative approximation for likelihoods \citep{Sellentin2014,Sarcevic2026}, will be considered in future work.

\subsection{Implementation} \label{subsec_implementation}
The previously discussed detector responses and the likelihood are implemented in \texttt{pespace} based on \texttt{taichi-lang} which is a domain-specific language with the \texttt{Python} frontend designed for high-performance numerical computation \citep{Hu2019,Hu2019a} and can use the just-in-time compilation to translate computationally intensive codes into optimized hardware instructions with various optimization techniques. 
Additionally, \texttt{taichi-lang} supports the data-oriented programming paradigm by introducing a flexible and efficient data container that can conveniently manipulate the layout of data in memory to improve the performance.
Furthermore, \texttt{taichi-lang} is backend neutral. Besides CPU, codes implemented with \texttt{taichi-lang} can be run on various GPU backends such as \texttt{CUDA} and \texttt{ROCm}.

{As mentioned in Sec.~\ref{subsec_response}, the time-to-frequency correspondence $t_f$ shown in Eq.~\ref{eq_tf} is required to compute the frequency domain responses using the method given by \citep{Marsat2018}.
For the robustness of fits, phenomenological waveform models commonly construct the phase by primarily fitting its derivative, and the ansatz is chosen for ${\mathrm{d} \Phi(f)}/{\mathrm{d} f}$. Thus, $t_f$ can be obtained directly from the ansatz coefficients rather than computing the phase derivative numerically.
However, the library \texttt{lalsimulation} provides waveform polarizations $\tilde{h}_+,\ \tilde{h}_\times$ as outputs, and does not offer an interface to directly access $t_f$.
Furthermore, to enable hardware acceleration and automatic differentiation through the entire computation of likelihood from a given set of parameters, waveform generation also needs to be incorporated into the \texttt{taichi-lang} scope.
Therefore, we reimplement the waveform models \texttt{PhenomXAS} \citep{Pratten2020} and \texttt{PhenomXHM} \citep{GarciaQuiros2020} in the separate package \texttt{tiwave}, to fully leverage the advanced features of \texttt{taichi-lang}, while providing $t_f$ obtained directly from the analytic expressions of the ansatz.
}

\begin{figure}
    \centering
    \includegraphics[width=0.8\columnwidth]{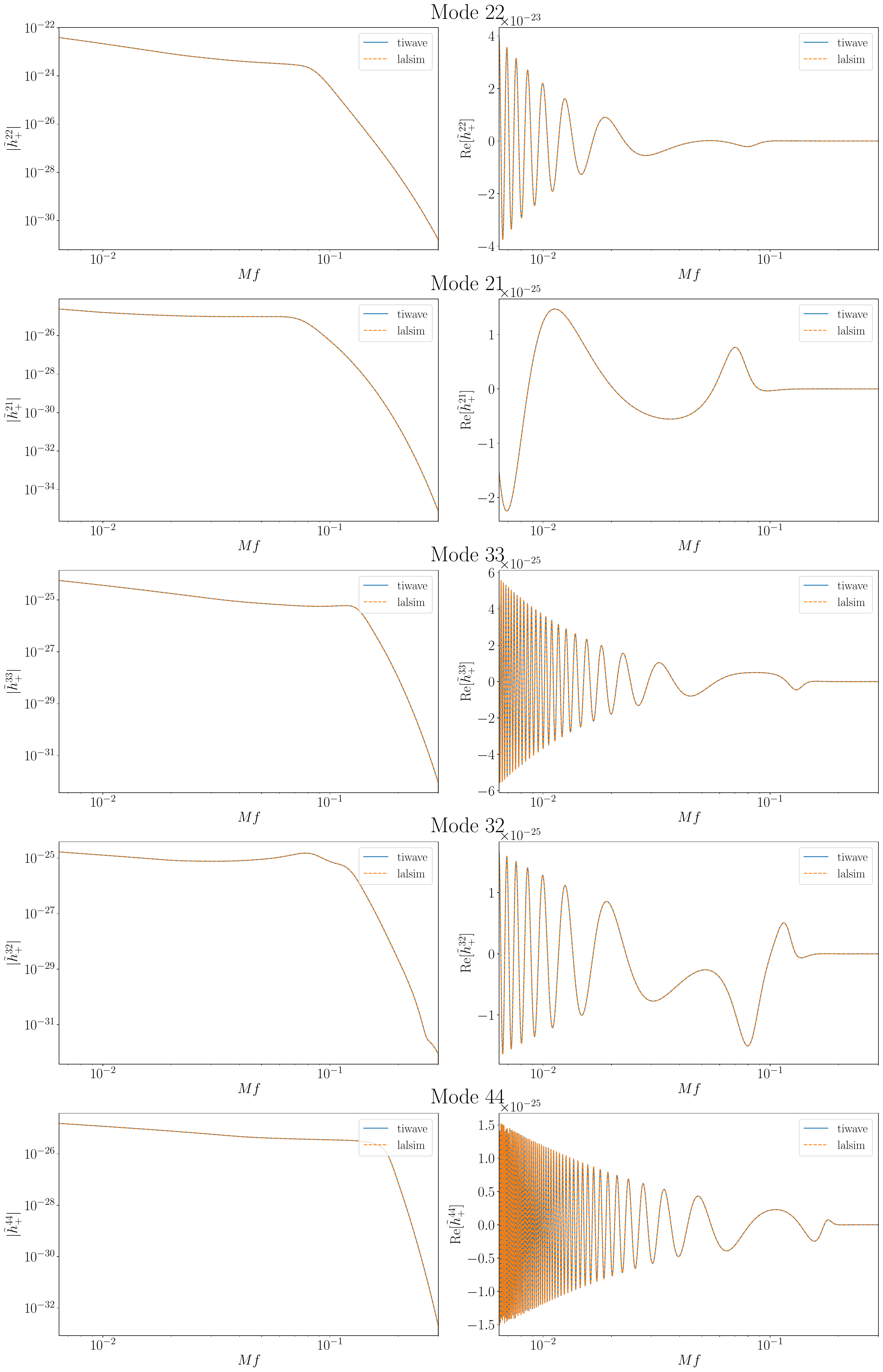}
    \caption{\textbf{Comparison of waveforms \texttt{PhenomXHM} generated by \texttt{lalsimulation} and \texttt{tiwave} for an example signal.}
    The left panels show the amplitude of waveforms, and the right panels show the real part of waveforms. The blue solid lines and orange dashed lines are used to represent waveforms from \texttt{tiwave} and \texttt{lalsimulation}, respectively.}
    \label{fig_wf_comp}
\end{figure}

Thorough tests are performed to ensure the reimplemented waveforms are consistent with waveforms from \texttt{lalsimulation}.
We check the ansatz coefficients using several representative examples, and compute the mismatch with waveforms from \texttt{lalsimulation} over the entire parameter space allowed by the waveform model.
The discrepancies are within acceptable tolerance. The maximum mismatch is of order $\mathcal{O}(10^{-15})$ except the case of mode 32 in \texttt{PhenomXHM}.
However, the large mismatch usually occurs in the case of extreme mass ratio and high spins. In the parameter space of $q<8$ and $|\chi_i|<0.8$ which is the valid regime of the numerical relativity hybrid surrogate model \texttt{NRHybSur3dq8} \citep{Varma2019b}, we find that the mismatches of mode 32 in our reimplemented waveform and in \texttt{lalsimulation} against \texttt{NRHybSur3dq8} are comparable.
Thus, we believe that at least for moderate mass ratio and spins, our reimplemented waveforms are still safe to use. More details are discussed in App.~\ref{app_mismatch_32}.
Additionally, we also find that a minor modification in waveform \texttt{PhenomXAS} can lead to a slight improvement in the mismatch against \texttt{NRHybSur3dq8}. Further details are presented in App.~\ref{app_mod_xas}.
In the main text, we adopt the waveform identical to that in \texttt{lalsimulation}.
In Fig.~\ref{fig_wf_comp}, we present a comparison of each mode in our reimplemented \texttt{PhenomXHM} and the original implementation in \texttt{lalsimulation} using an example signal.

\begin{sidewaysfigure}
    \centering
    \includegraphics[width=\linewidth]{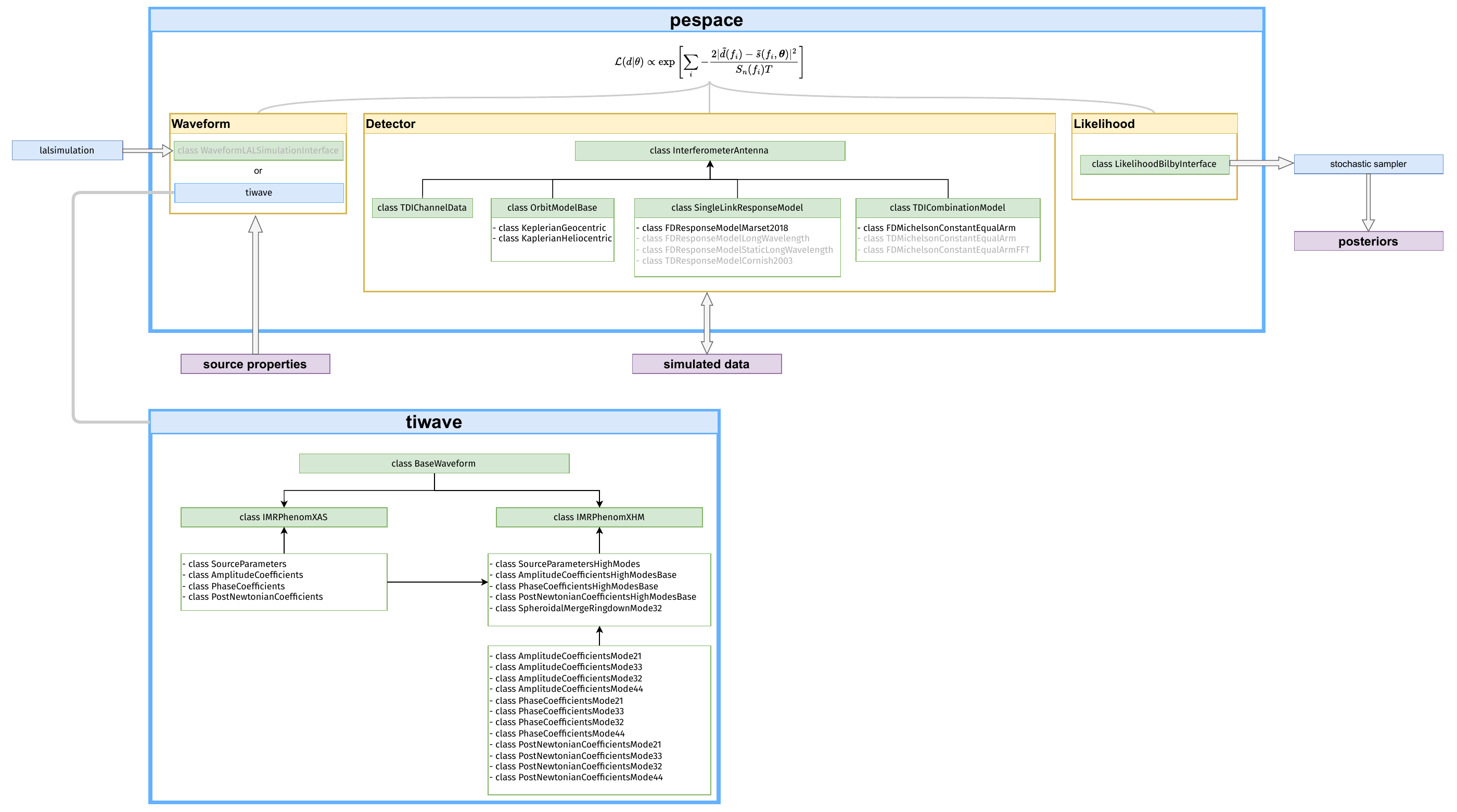}
    \caption{\textbf{A illustration for main modules in \texttt{pespace} and \texttt{tiwave}.}
    Blue boxes denote packages, green boxes denote involved modules, and purple boxes denote input or output parameters or data. Modules shown in gray indicate components that are still under development or not used in this paper.}
    \label{fig_modules}
\end{sidewaysfigure}

{An overview of main modules in \texttt{pespace} and \texttt{tiwave} is shown in Fig.~\ref{fig_modules}, where blue boxes denote packages, green boxes denote involved modules, and purple boxes denote input or output parameters or data. Modules shown in gray indicate components that are still under development or not used in this paper.
\texttt{tiwave} is responsible for waveform generation, supports generating waveforms of \texttt{PhenomXAS} and \texttt{PhenomXHM} with high performance.
$t_f$ is generated directly from ansatz coefficients, except for the merge-ringdown range of mode 32 in \texttt{PhenomXHM} where the ansatz is constructed in the spheroidal-harmonic basis. $t_f$ for the phase in the spherical-harmonic basis is obtained numerically.
\texttt{pespace} is used to compute detector responses from GW signals, evaluate the likelihood function, and communicate with external sampling algorithms. The main functionality is centered on the likelihood function and can be divided into three components: waveform access, detector response computation, and likelihood evaluation. The corresponding modules are organized into yellow boxes in Fig.~\ref{fig_modules}.
For waveform access, in addition to \texttt{tiwave}, we also provide an interface to access a wide variety of waveform models in \texttt{lalsimulation}.
However, using this interface requires additional overhead of converting \texttt{numpy.array} to \texttt{taichi.field}, which thus degrades the performance.
Moreover, as mentioned previously, \texttt{lalsimulation} does not provide access to $t_f$. We simply use the post-Newtonian phase to obtain the time-to-frequency correspondence for the inspiral range, and set time to coalescence time for merge-ringdown range considering that the orbital evolution of detectors over this period is relatively negligible. But the validity of this approach has yet to be strictly verified.
Therefore, in this work, we only use the waveforms \texttt{PhenomXAS} and \texttt{PhenomXHM} in \texttt{tiwave}.
The detector component primarily handles tasks including storing and processing simulated observational data, computing single link responses, and combining them with TDI. Different orbit models are used to define different detectors.
The likelihood component is used to compute the likelihood function and communicate with external stochastic sampling algorithms.
Currently, we rely on the unified interface provided by \texttt{bilby} \citep{Ashton2019} to support multiple external samplers.
}

Automatic differentiation provides an efficient and accurate approach for computing derivatives or gradients that are essential for the computation of the Fisher matrix or gradient-based stochastic sampling algorithms.
Automatic differentiation evaluates derivatives by applying the chain rule to elementary operations, and propagates derivatives along the chain to obtain the derivatives of outputs with respect to inputs \citep{Corliss2006,Baydin2015,Margossian2019}.
The propagation has two different modes, the forward mode and the backward mode. The forward mode propagates alongside the function computation, and is more efficient for cases of small inputs and large outputs such as computing the Fisher matrix where the partial derivatives of detector responses with respect to source parameters are required.
The backward mode propagates derivatives in reverse from given outputs, and is more suitable for cases of large inputs and small output such as computing the gradients of the likelihood function.

\begin{figure}
    \centering
    \includegraphics[width=0.8\columnwidth]{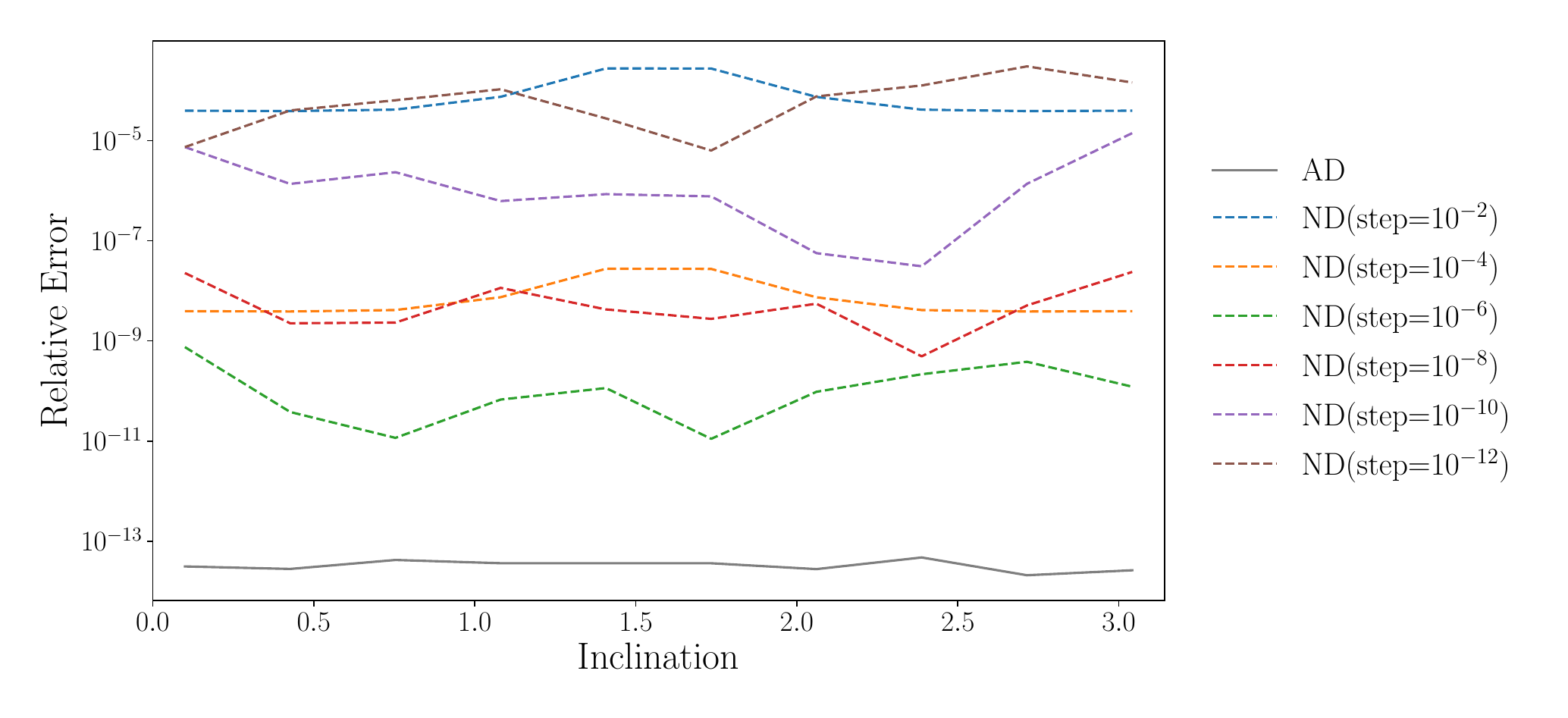}
    \caption{\textbf{Comparison of numerical errors of derivatives obtained by automatic differentiation (AD) and numerical differentiation (ND). }
    We use the derivative of the optimal SNR $\rho$, where $\rho^2\equiv\langle s|s\rangle$, with respect to the inclination $\iota$, which can be computed analytically, to illustrate the accuracy of derivative computation.
    The vertical axis shows the relative difference between symbolic derivatives (SD) and derivatives obtained from automatic or numerical differentiation, which is given by $\mathrm{abs}\left[ (\frac{\partial \rho}{\partial \iota}|_{\rm SD} - \frac{\partial \rho}{\partial \iota}|_{\rm AD, or ND})/{\frac{\partial \rho}{\partial \iota}|_{\rm SD}}\right]$.
    The results of numerical differentiation are obtained by the central finite difference scheme with different steps indicated by dashed lines in different colors.
    As discussed in Sec.~\ref{subsec_implementation}, too small (or large) steps can induce significant round-off (or truncation) errors.
    By contrast, automatic differentiation can avoid manually tuning step sizes, and offer accurate computation of derivatives, as shown by the gray line in the above figure.}
    \label{fig_nerr_diota}
\end{figure}

One of the key advantages of automatic differentiation is the numerical stability compared to the finite difference method. 
In the finite difference method, the choice of the step size is usually delicate and plays a critical role in the accuracy of the obtained derivatives.
If the step is too large, the obtained difference of the function values between the two points cannot accurately represent the local features of the function. The truncation errors become significant, thereby degrading the accuracy of derivatives.
In contrast, if the step size is too small, the phenomenon of catastrophic cancellation can lead to unreliable results for subtraction between two nearby values. The round-off errors may be dominant in the obtained difference.
This can be clearly shown in Fig.~\ref{fig_nerr_diota} where we use the example of the derivative of the optimal SNR with respect to the inclination, which can be obtained by symbolic differentiation, to illustrate the errors of automatic differentiation and numerical differentiation with different steps.
It can be seen that the errors of numerical differentiation can be reduced by using smaller steps, but increase when the step is too small.
By contrast, the derivatives from automatic differentiation are much closer to results from symbolic differentiation, indicating that automatic differentiation can provide accurate derivatives approaching the machine precision limit.

Manually tuning the steps of each parameter is usually cumbersome and nontrivial in computing the Fisher matrix using numerical differentiation.
Automatic differentiation can avoid this procedure.
By leveraging the differentiable programming capability of \texttt{taichi-lang}, we can compute the gradients of likelihood or the partial derivatives of detector responses more accurately and efficiently.
Currently, the detector response implemented in \texttt{pespace} supports two modes of automatic differentiation, but in \texttt{tiwave}, automatic differentiation is available only for \texttt{PhenomXAS} and is limited to the backward mode.

\section{Results} \label{sec_results}
{Compared with previous tools \citep{Cornish2020a,Marsat2021,Katz2020}, new features offered by \texttt{pespace} include: enabling hardware acceleration and automatic differentiation; incorporating detectors of LISA, Taiji, Tianqin and allowing parameter estimation for joint observations by the detector network; supporting more recent phenomenological waveform models \texttt{PhenomXAS} and \texttt{PhenomXHM}.
We demonstrate these features by performance tests, full Bayesian parameter estimation of a typical MBHB signal, and the Fisher matrix analysis for a subset of extrinsic parameters in this section.

Hardware acceleration allows computations to be completed in significantly reduced time. We compare the computational cost of waveform generation and likelihood evaluation for CPU-only and GPU-accelerated executions in Sec.~\ref{subsec_perf}.
Sec.~\ref{subsec_pe_results} shows the improvement of parameter measurement by incorporating multiple detectors and using the waveform with higher modes \texttt{PhenomXHM} in the full Bayesian parameter estimation.
Automatic differentiation offers efficient and accurate computations of derivatives without manually tuning step sizes in the finite difference method. In Sec.~\ref{subsec_fisher}, we compare the approximated posteriors obtained by the Fisher matrix based on automatic differentiation and posteriors from Bayesian parameter estimation with the noise-free likelihood function. 

However, in this paper we do not provide a comparison of the total time required to complete the full Bayesian parameter estimation with previous tools. Because the overall runtime strongly depends on the sampling algorithm. 
Automatic differentiation in \texttt{pespace} is enabled by \texttt{taichi-lang} which is a relatively recent high-performance computational framework. A wide variety of existing gradient-based samplers in the \texttt{jax} ecosystem cannot be directly used. Currently, we rely on the nested sampler \texttt{multinest} \citep{Feroz2007,Feroz2008,Feroz2013} to explore the parameter space, where the gradient information of likelihood surface is not used.
In future work, we will improve the sampling process to fully leverage advantages of automatic differentiation provided by \texttt{taichi-lang}.
A detailed comparison of overall runtime of the entire Bayesian parameter estimation will be presented in future work.
}

\subsection{Performance} \label{subsec_perf}
\begin{figure}
    \centering
    \includegraphics[width=0.8\columnwidth]{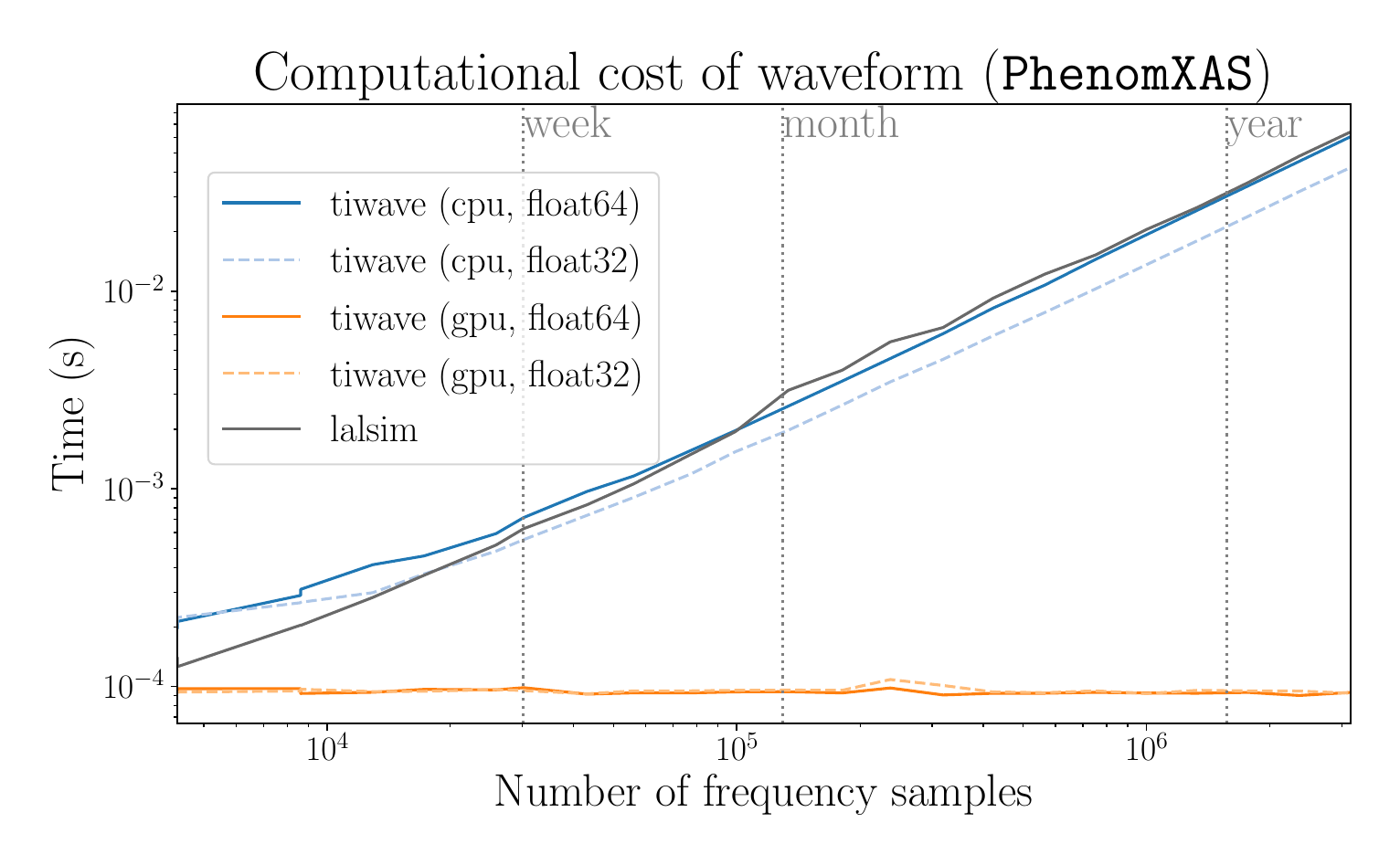}
    \includegraphics[width=0.8\columnwidth]{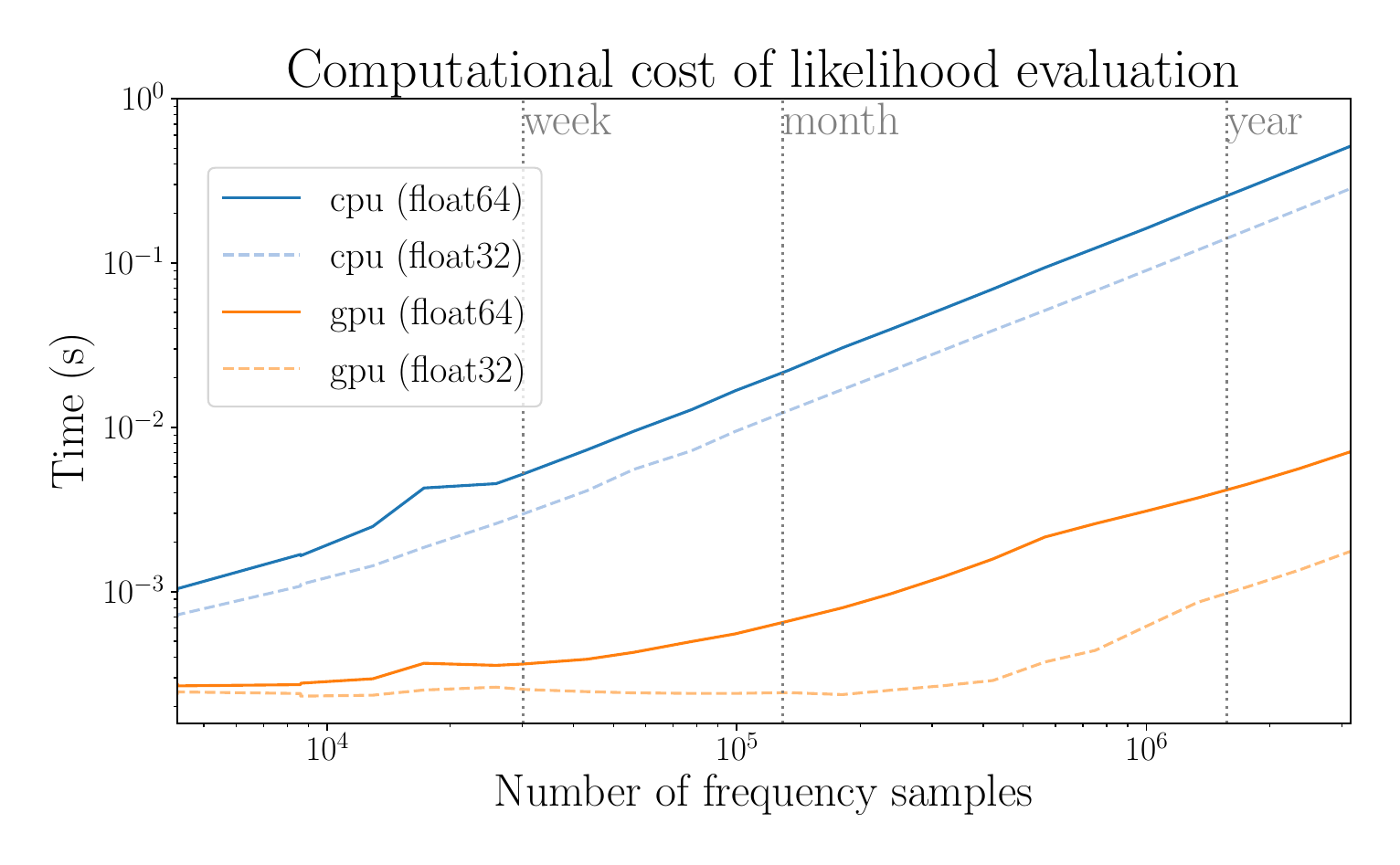}
    \caption{\textbf{Computational cost of waveform generation and likelihood evaluation.} 
    The blue and orange lines denote results of tests performed using CPU only and using GPU acceleration, respectively, and the solid and dashed lines are used to represent the different settings of floating-point precision.
    The vertical dotted lines denote the corresponding time domain duration with the sampling interval of 10 seconds.
    The number of frequency samples is determined by the length of time samples through \texttt{numpy.fft.rfftfreq} with the truncation of the range $[10^{-4}, 5\times10^{-2}]$.
    The computational times are obtained by averaging over 10 independent runs with parameters randomly sampled in the parameter space.
    These tests are performed on a computing platform equipped with a CPU of AMD Threadripper 7955WX and a GPU of NVIDIA RTX 5000 Ada Generation. The CPU is forced to use only one core in tests.
    Computational cost of waveform generation using \texttt{lalsimulation} is also shown in the upper panel, where we use the interface \texttt{SimInspiralChooseFDWaveformSequence} to get waveforms on specific frequency grids, all waveform flags use default settings except that the multibanding is switched off, and \texttt{OpenMP} parallelization is disabled.
    The likelihood evaluation incorporates one LISA-like detector with three orthogonal TDI channels.}
    \label{fig_cost}
\end{figure}

First, we perform the tests of computational cost of waveform generation and likelihood evaluation, and results are shown in Fig.~\ref{fig_cost}. 
We compare the performance of computations executed solely on CPU and with GPU acceleration. Details of the configuration for the tests are given in the caption of Fig.~\ref{fig_cost}, and we emphasize that the performance strongly depends on the hardware and software environment. The timing results may vary across different platforms.
The results demonstrate that a substantial speedup can be achieved with GPU acceleration. For signals with durations of several months, GPU acceleration can reduce the computational time by two orders of magnitude.
In the performance test of waveform generation, we also measure the cost of implementation provided in \texttt{lalsimulation} (with multibanding \citep{GarciaQuiros2020a} and \texttt{OpenMP} parallelization disabled) as a reference for comparison.
When executed solely on CPU, our implementation can offer comparable performance to the native \texttt{C} implementation of \texttt{lalsimulation}, while retaining the pythonic readability and usability.
If low floating-point precision is adopted, which might be useful for preliminary online searches or for initializing Bayesian parameter estimation with a coarse maximum likelihood estimation, the computational cost can be further reduced.

It is observed that when executing with GPU acceleration, the computation time remains nearly constant as the number of frequency samples increases in the situation of waveform generation and the situation of likelihood evaluation using \texttt{float32} for modest sample numbers.
This behavior is expected and does not indicate any error, as the computation is dominated by fixed overheads like kernel launch, task dispatch, memory management, etc. The computation saturates the available parallelism, and increasing frequency samples within the parallelism limit does not introduce additional serial work \citep{Cheng2014}. 

\subsection{Full Bayesian parameter estimation} \label{subsec_pe_results}
To demonstrate the basic functionality of \texttt{pespace}, we perform the full Bayesian parameter estimation using a typical MBHB signal.
Parameters of the injected signal are shown in Tab.~\ref{tab_pe_res}.
We use the 2.0-generation TDI combination to compute detector responses and incorporate $(A,\ E,\ T)$ channels in likelihood evaluation.
Simulated data for analyses are generated on the frequency grid corresponding to the time series with the duration of 655360 seconds ($\sim 7.6$ days) and the sampling interval of 10 seconds. As discussed in \citep{Cornish2020a}, this duration is sufficient to capture the majority of the SNR of the injected signals.

For the noise model, we consider the ideal stationary Gaussian noise as discussed in Sec.~\ref{subsec_pe}, and the noise properties are assumed to be known when performing parameter estimation of the GW source.
The noise budgets used for LISA, Taiji, and Tianqin are from the references \citep{Babak2021,Ruan2020a,Li2023}, respectively.
We only consider the components of the acceleration noise and the optical metrology system noise.
The confusion noise \citep{Belczynski2010,Ruiter2010,Nelemans2001,Niu2024a} from the unresolvable foreground Galactic compact binaries is left to future work.
The confusion noise has the feature of nonstationarity owing to the orbital motion of detectors, which may invalidate the underlying assumption of Eq.~\ref{eq_diag_Sigma_to_Sn} and lead to the likelihood function Eq.~\ref{eq_ll_fd} being no longer applicable.
We will consider the confusion noise in separate modules within the global-fitting framework in future work to address the influence of the nonstationarity and to extract the population properties of Galactic compact binaries.

We consider three distinct scenarios for parameter estimation. First, we perform parameter estimation using the waveform \texttt{PhenomXAS} and the detector LISA.
It has been pointed out that the subdominant multipoles of gravitational radiations, i.e., higher modes, play a crucial role in breaking degeneracies \citep{Marsat2021}, therefore, we use the waveform \texttt{PhenomXHM} where the higher modes 21, 33, 32, 44 are included in the second case.
In the most optimistic scenario, the proposed space-borne GW detectors Taiji, LISA, and Tianqin may operate simultaneously in the future and have overlapping observing periods.
The same GW signal can be jointly observed by the network of three detectors. 
Accordingly, in the third scenario, we consider this optimistic case by including all three detectors in the parameter estimation and incorporating high modes in the waveform.
We implement different detectors in the code by adopting different orbits of the constellation.
Currently, we use the the analytical heliocentric orbits for LISA and Taiji \citep{Stas2020,Ren2023}, and the analytical geocentric orbit for Tianqin \citep{Hu2018}. We impose initial values of $-20^\circ$, $0^\circ$, and $20^\circ$ for the revolution of the constellations around the Sun to specify the relative positions of different detectors.
Support for high-precision numerical orbits will be extended in future work.

\begin{figure}
    \centering
    \includegraphics[width=\columnwidth]{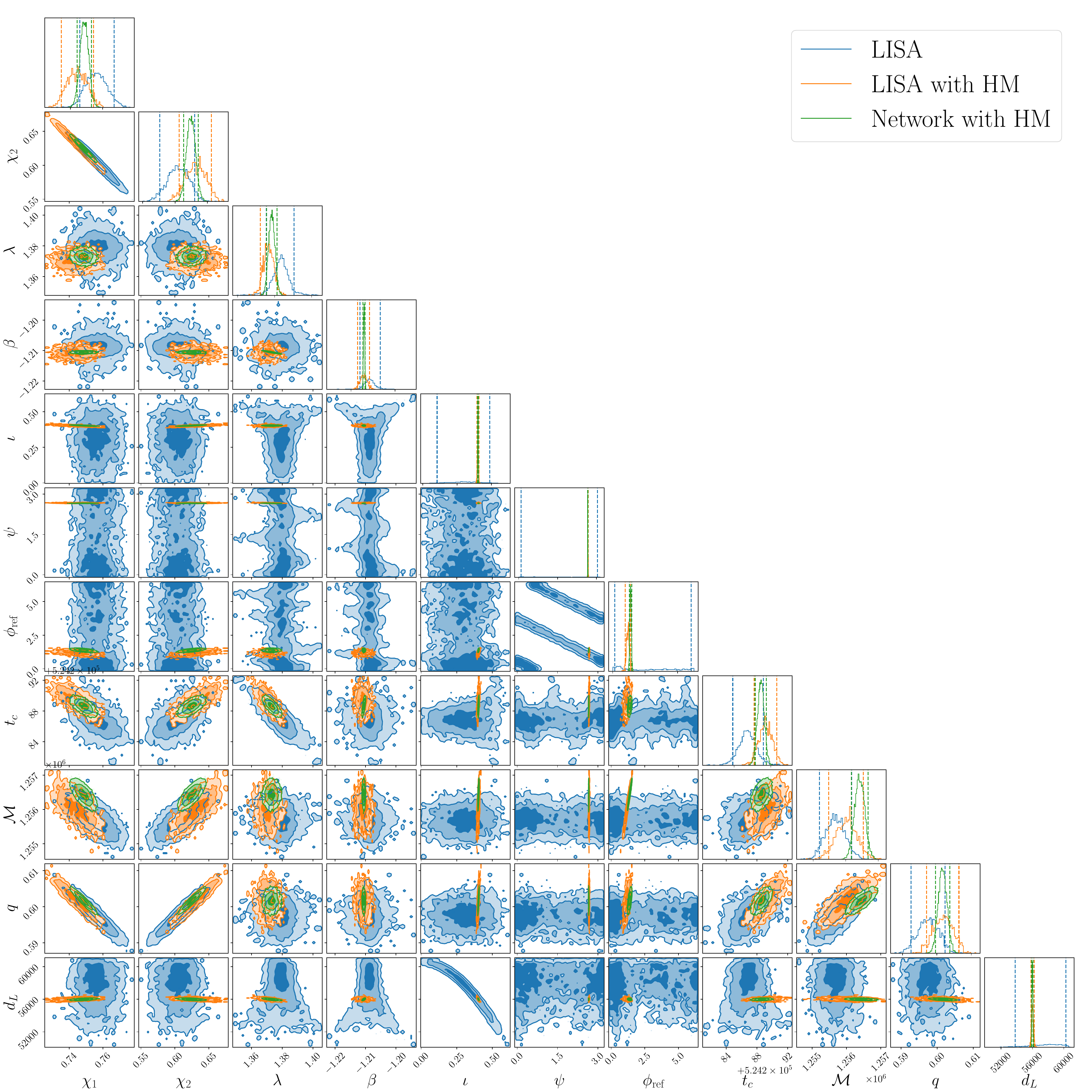}
    \caption{\textbf{Posteriors obtained from the full Bayesian parameter estimation.}
    The blue, orange, and green colors denote results of three scenarios: the single LISA detector observation and using the waveform with only the dominant 22 mode; the single LISA detector observation and using the waveform with higher modes (HM) including 21, 33, 32, 44 modes; the joint observation by the LISA-Taiji-Tianqin detector network and using the waveform with higher modes.
    The dashed lines in marginalized posterior histograms mark the $90\%$ credible interval, defined by the 5th and 95th percentiles.
    }
    \label{fig_corner_pe}
\end{figure}

The obtained posteriors are shown in Fig.~\ref{fig_corner_pe}, and the recovered values of source parameters and their $90\%$ credible intervals are summarized in Tab.~\ref{tab_pe_res}.
It is evident that including higher modes can significantly improve the measurement accuracy of inclination, luminosity distance, polarization, and reference phase.
Owing to distinct dependencies on different harmonics, including higher modes can effectively break the well-known degeneracy between inclination and luminosity distance.
For the example signal, we observed that the measurement uncertainty in inclination can be improved by nearly two orders of magnitude, and the uncertainty in luminosity distance can be improved by one order of magnitude by using the waveform incorporating higher modes 21, 33, 32, and 44.
Furthermore, the polarization and reference phase can hardly be effectively constrained by the waveform with only the 22 mode.
By contrast, after considering higher modes, the polarization can be measured within the range of $0.03$ rad, and the reference phase can be constrained within $0.4$ rad, at the $90\%$ credible interval.
We can also observe from Tab.~\ref{tab_pe_res} that the measurement accuracy of extrinsic parameters $\lambda$, $\beta$, and $t_c$ show a modest improvement after including higher modes in the waveform.
In the most optimistic scenario, if the signal is observed jointly by all three detectors, the measurement accuracy of all parameters can be enhanced by approximately a factor of two overall, with the uncertainty in $\beta$ and $\psi$ in particular being reduced by nearly an order of magnitude.

\subsection{The Fisher matrix based on automatic differentiation}\label{subsec_fisher}
As discussed in Sec.~\ref{subsec_implementation}, automatic differentiation can offer efficient and accurate computation of derivatives up to the limit of machine precision, which can circumvent the need of manually tuning step sizes to trade off the truncation errors and the round-off errors in the computation of Fisher matrix using the finite difference method.
Here, we use a subset of extrinsic parameters including $\lambda$, $\beta$, $\psi$, and $t_c$ to demonstrate the approximated posteriors given by the Fisher matrix based on automatic differentiation.
The fiducial parameters are the same as the injected signal used in the full Bayesian parameter estimation shown in Tab.~\ref{tab_pe_res}, and configurations like duration, sampling interval, TDI channels, noise features, etc.~are the same as those used in the last subsection.
In this analysis, we only use the detector of LISA, and only consider the parameters related to detector responses. The parameters related to waveform generation are fixed.
Since the derivatives of detector responses with respect to source properties rely on forward mode automatic differentiation, which currently is not fully supported by \texttt{tiwave}, discussions of parameters related to waveform generation are left to future work.

\begin{figure}
    \centering
    \includegraphics[width=0.5\columnwidth]{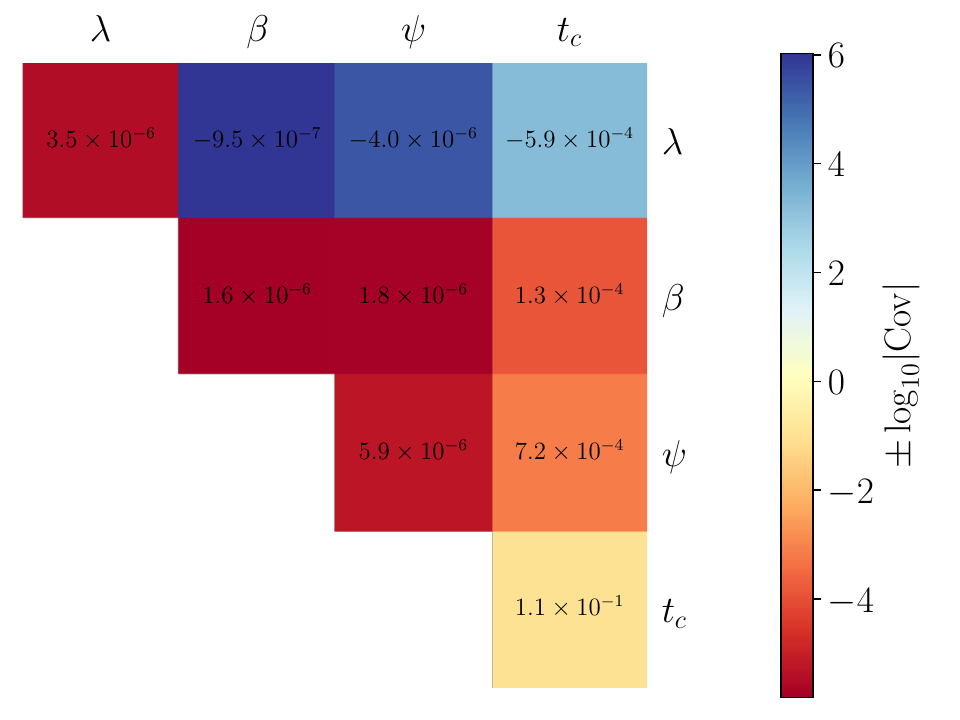}
    \caption{\textbf{Heatmap of the covariance matrix obtained from the Fisher matrix computed via automatic differentiation.} 
    Warm colors indicate positive correlations and cool colors indicate negative correlations for the off-diagonal elements. 
    Since a logarithmic color scale is used, the sign convention in the colorbar appears inverted.
    }
    \label{fig_cov}
\end{figure}

\begin{figure}
    \centering
    \includegraphics[width=0.8\columnwidth]{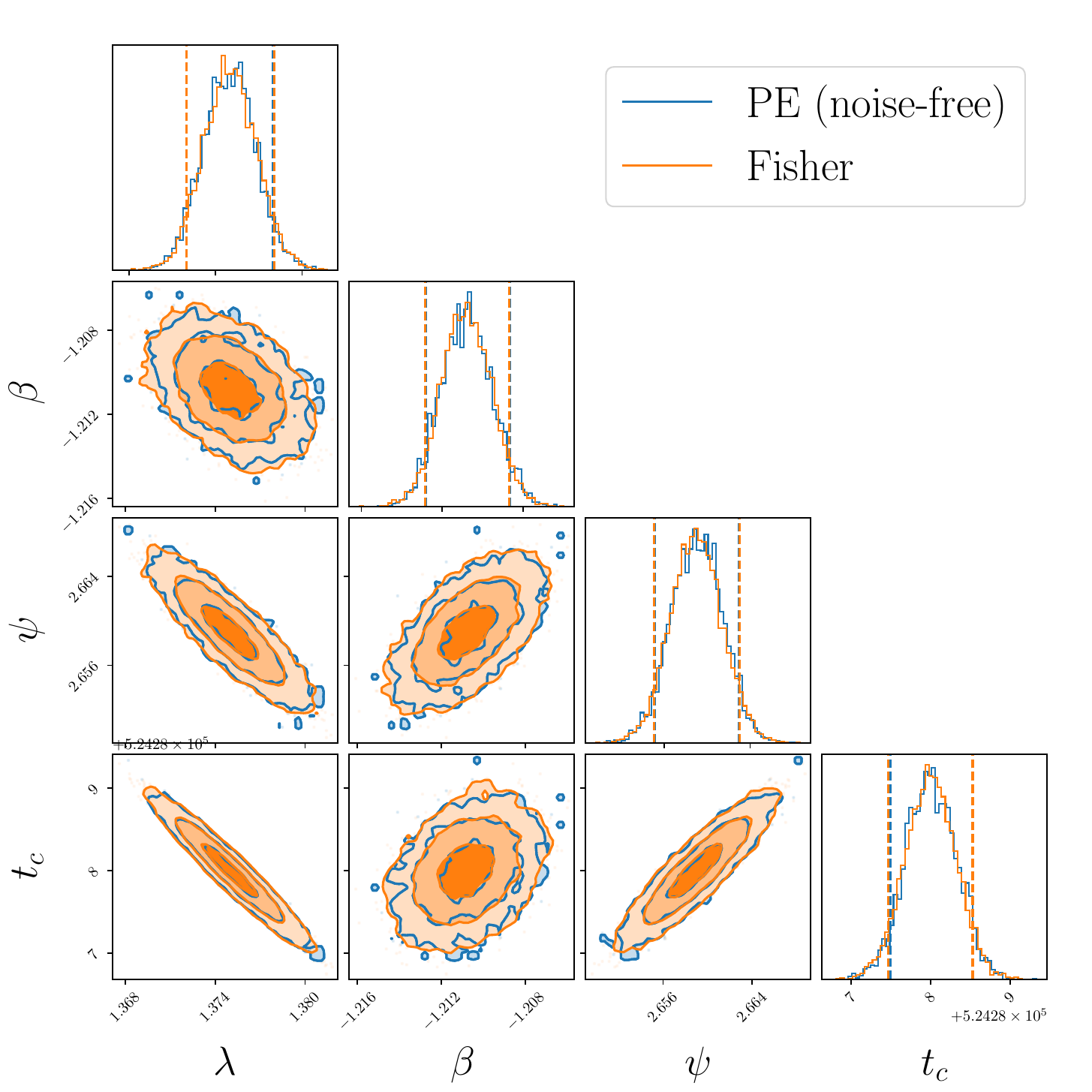}
    \caption{\textbf{Approximated posteriors given by the Fisher matrix and posteriors from the Bayesian parameter estimation}.
    The blue and orange colors denote results obtained by parameter estimation with stochastic sampling and the Fisher matrix, respectively.
    The dashed lines in marginalized probability histograms indicate the $5\%$ and $95\%$ quantiles.
    We only consider a subset of extrinsic parameters $(\lambda,\ \beta,\ \psi,\ t_c)$, and fix parameters related to waveform generation at the fiducial values.
    To avoid biases induced by fluctuations of noise realizations, we adopt the noise-free likelihood function in the parameter estimation.
    We emphasize that this result is mainly used to illustrate the agreement of posteriors approximated by the Fisher matrix and posteriors actually sampled in the Bayesian parameter estimation, uncertainties of parameters may be optimistically estimated, owing to not all parameters are varied in the analysis.
    }
    \label{fig_corner_fisher}
\end{figure}

As reviewed in Sec.~\ref{subsec_pe}, from the Bayesian perspective and considering uniform priors, the inverse of the Fisher matrix can be interpreted as the covariance matrix for the posteriors. 
The resulting covariance matrix for the parameters $(\lambda,\ \beta,\ \psi,\ t_c)$ is visualized as a heatmap in Fig.~\ref{fig_cov}.
The covariance matrix quantifies the uncertainty and correlations of the estimated parameters, where the diagonal elements represent the variances of individual parameters, while the off-diagonal elements encode the covariances between pairs of parameters and measure the degree to which uncertainties in different parameters are correlated.
From the Bayesian point of view, the posterior distribution is approximated as a multivariate Gaussian distribution with the fiducial parameters as means and the inverse of the Fisher matrix as the covariance.
Thus, we can make a comparison between this approximated posterior distribution and that from full Bayesian parameter estimation with stochastic sampling.
The obtained results are shown in Fig.~\ref{fig_corner_fisher}.
As discussed in \citep{Iacovelli2022,Rodriguez2013a}, due to contributions of noise realizations, the posteriors might not peak exactly at the injected values, while the expected value of biases from repeated experiments with different noise realizations is zero.
Therefore, when performing the comparison, to avoid fluctuations induced by specific noise realizations, we adopt the noise-free likelihood in parameter estimation where the log-likelihood function takes the form of $-\frac12 \langle s(\boldsymbol{\theta}) - s(\boldsymbol{\theta}_0) | s(\boldsymbol{\theta}) - s(\boldsymbol{\theta}_0) \rangle$ \citep{Marsat2021}.
The approximated posteriors sampled from the multivariate Gaussian distribution given by the Fisher matrix and the posteriors obtained by the full Bayesian parameter estimation with stochastic sampling are presented in Fig.~\ref{fig_corner_fisher}, where posterior samples from the two methods show good agreement.
However, we emphasize that the analysis is restricted to a subset of extrinsic parameters. The parameters related to waveform generation are fixed at their injected values, and not varied in parameter estimation.
The results shown in Fig.~\ref{fig_corner_fisher} are mainly used to illustrate the agreement of posteriors obtained by the Fisher matrix and Bayesian parameter estimation, uncertainties of these parameters might be underestimated.

\begin{sidewaystable}
\centering
    \begin{tabular}{llllll}
        \toprule
        \textbf{Parameters} & \textbf{Injected} & \textbf{Prior} & \textbf{Recovered} & \textbf{Recovered (HM)} & \textbf{Recovered (HM, network)} \\
        \midrule
        Chirp mass, $\mathcal{M}$ ($10^6 M_\odot$) & 1.256227 & Uniform[0.5, 2.0]        & $1.255683^{+0.000501}_{-0.000493}$ & $1.256027^{+0000517}_{-0.000550}$  & $1.256429^{+0.000258}_{-0.000251}$  \\
        Mass ratio, $q$                            & 0.6      & Uniform[0.05, 0.99]      & $0.59752^{+0.00486}_{-0.00503}$    & $0.60190^{+0.00451}_{-0.00492}$    & $0.6016^{+0.00211}_{-0.00207}$      \\
        Aligned spin of primary, $\chi_1$          & 0.75     & Uniform[-0.99, 0.99]     & $0.7560^{+0.0110}_{-0.0102}$       & $0.74420^{+0.0102}_{-0.00984}$     & $0.7486^{+0.00436}_{-0.00452}$      \\
        Aligned spin of secondary, $\chi_2$        & 0.62     & Uniform[-0.99, 0.99]     & $0.6047^{+0.0257}_{-0.0282}$       & $0.6316^{+0.0246}_{-0.0253}$       & $0.6245^{+0.0115}_{-0.0112}$        \\
        Luminosity distance, $d_L$ (Gpc)           & 56       & LogUniform[10, 5000]     & $57.86^{+2.42}_{-4.06}$            & $56.017^{+0.205}_{-0.213}$         & $55.995^{+0.101}_{-0.101}$          \\
        Inclination, $\iota$                       & 0.4      & Sine$[0, \pi]$           & $0.302^{+0.185}_{-0.198}$          & $0.40138^{+0.00728}_{-0.00739}$    & $0.40237^{+0.00318}_{-0.00316}$     \\
        Reference phase, $\phi_{\mathrm{ref}}$     & 1.3      & Uniform$[0, 2\pi]$       & $2.91^{+3.14}_{-2.76}$             & $1.182^{+0.192}_{-0.209}$          & $1.3786^{+0.0978}_{-0.0964}$        \\
        Ecliptic longitude, $\lambda$              & 1.375    & Uniform$[0, 2\pi]$       & $1.37922^{+0.00886}_{-0.00975}$    & $1.37087^{+0.00572}_{-0.00557}$    & $1.37306^{+0.00357}_{-0.00345}$     \\
        Ecliptic latitude, $\beta$                 & -1.2108  & Cosine$[-\pi/2, \pi/2]$  & $-1.20871^{+0.00363}_{-0.00329}$   & $-1.21074^{+0.00199}_{-0.00204}$   & $-1.210558^{+0.000338}_{-0.000345}$ \\
        Polarization, $\psi$                       & 2.659    & Uniform$[0, \pi]$        & $1.36^{+1.68}_{-1.27}$             & $2.6721^{+0.0149}_{-0.0166}$       & $2.66108^{+0.00620}_{-0.00629}$     \\
        Coalescence time, $t_c$ (s)                & 524288   & Uniform[524188, 524388]  & $524286.85^{+2.12}_{-2.03}$        & $524289.26^{+1.56}_{-1.55}$        & $524288.642^{+0.761}_{-0.781}$      \\
        \bottomrule
    \end{tabular}
    \caption{\textbf{Values of source parameters for the injected signal, priors used in parameter estimation, and recovered values of signal parameters along with their $90\%$ credible intervals.}}
    \label{tab_pe_res}
\end{sidewaystable}

\section{Summary}\label{sec_summary}
With the recent development of high-performance computation hardware, heterogeneous computing where synergies of CPUs, GPUs, and various specialized accelerators can significantly improve the computational performance is becoming increasingly prevalent in scientific computation.
As the volume and complexity of computational demand for data analysis of future space-borne GW observations rapidly growing, next-generation data analysis tools need to fully exploit contemporary high-performance computing hardware.
However, programming and optimizing on heterogeneous computing architectures are usually complicated, requiring expertise in low-level hardware-specific details, which limits their applications in GW data analysis.
In this context, emerging frameworks like \texttt{taichi-lang}, \texttt{jax}, etc.~that provide high-level abstractions, automatic parallelization, and backend-agnostic execution may offer promising solutions to lower barriers of utilizing contemporary high-performance computing hardware.
Furthermore, automatic differentiation enables efficient and accurate computations of derivatives or gradients, which are particularly valuable for optimization and parameter estimation tasks in GW data analysis.

In this work, we focus on the infrastructure development for data analysis of future space-borne GW missions, and introduce a new tool for detector response generation and likelihood evaluation with features of GPU acceleration and automatic differentiation.
The main improvements compared with previous similar tools \citep{Cornish2020a,Marsat2021,Katz2020} include: incorporating proposed detectors of LISA, Taiji, Tianqin and enabling parameter estimation with joint observations by the detector network; reimplementing waveform models \texttt{PhenomXAS} and \texttt{PhenomXHM} using \texttt{taichi-lang} to enable improved performance with GPU acceleration; supporting automatic differentiation for efficient and accurate computation of derivatives or gradients which can be used in the Fisher matrix analysis or gradient-based stochastic sampling algorithms.

We perform tests on the performance of waveform generation and likelihood evaluation at individual samples in the parameter space.
In CPU-only execution, the reimplemented waveform \texttt{PhenomXAS} using \texttt{taichi-lang} has comparable performance to the \texttt{C} implementation in \texttt{lalsimulation}.
In the case of GPU acceleration, the waveform generation and likelihood evaluation can be accelerated by nearly two orders of magnitude for signals of several months on the computing platform used for the test.
Using a typical MBHB signal, we perform the full Bayesian parameter estimation with three different scenarios. First, we consider the observation by the single LISA detector and use \texttt{PhenomXAS} where only the dominant 22 mode is included as the waveform model.
Next, we incorporate higher modes 21, 33, 32, 44 in the waveform model to perform parameter estimation.
Finally, we consider the joint observation by the detector network including LISA, Taiji, and Tianqin.
The results demonstrate that parameter measurements of inclination, luminosity distance, polarization, and reference phase can be significantly improved by including higher modes.
Joint observations by the detector network can reduce the parameter measurement uncertainties by approximately a factor of two, and in particular the measurement accuracy of ecliptic latitude and polarization can be enhanced by nearly an order of magnitude.
Powered by the differentiable programming in \texttt{taichi-lang}, we compute the Fisher matrix for a subset of extrinsic parameters and make the comparison between the approximated posteriors from the Fisher matrix analysis and the posteriors actually sampled from the parameter space with the noise-free likelihood function, which shows good agreement. 

To address real data from future detectors, in future work we will continue to improve our tool in three main aspects. 
First, data from space-borne detectors are signal-dominated, where a massive number of different types of GW signals are tangled together. The global-fitting framework is required to extract properties of overlapped signals. However, since the correlations among different types of signals are moderate, in practice the Gibbs update scheme, where stochastic sampling is performed iteratively for each source type in a cyclic manner, is adequate to explore extremely expansive parameter space of the global-fitting. The current module for MBHB signals needs to be integrated into the global-fitting framework.
Next, in real observed data, various imperfections like glitches, data gap, etc.~will inevitably be present. The ideal stationary Gaussian noise model may be insufficient to characterize the noise behavior. More realistic noise models are required for accurate parameter estimation. Furthermore, the noise properties are not known a priori, and it is unlikely to acquire segments of pure noise in signal-dominated data, which prevents the applicability of methods like Welch averaging for analysis of off-source noise behavior. Thus, the noise characteristics may also need to be inferred within the global-fitting framework simultaneously with properties of GW sources. 
Finally, the responses of space-borne detectors to GWs are more complex. The response model in frequency domain adopted in our tool has only been validated for signals chirping fast enough, and the applicability for effects like the double-spin precession still requires further investigation. Future work may require modeling jointly waveform templates and detector responses, or alternatively adopting the native time-domain response model.

Ultimately, we aim to provide a convenient and easy-to-use infrastructure to facilitate the development of data analysis methods and investigation of various scientific questions in preparatory science of future space-borne GW missions.

\appendix
\section{Numerical errors of mode 32} \label{app_mismatch_32}
To ensure the fidelity of the reimplemented waveforms in \texttt{tiwave}, we perform comprehensive verification by checking values of ansatz parameters for representative example signals and comparing mismatches with waveforms generated by \texttt{lalsimulation}.
The mismatch is used to quantify the difference between two waveforms and is defined as
\begin{equation} \label{eq_mismatch}
    \mathrm{Mismatch}(h_1, h_2) = 1-\max_{t_c, \phi_c} \left\langle \hat{h}_1 \middle| \hat{h}_2\right\rangle,
\end{equation}
where the inner product $\langle\cdot|\cdot\rangle$ is defined in Eq.~\ref{eq_inner_product}, and $\hat{h}_i$ is the waveform after normalization according to $\hat{h}_i = h_i/\sqrt{\langle h_i| h_i\rangle}$.
Time and phase shifts are adjusted to align two waveforms by using the optimization algorithm of dual annealing in \texttt{scipy}.

\begin{figure}
    \centering
    \includegraphics[width=0.8\columnwidth]{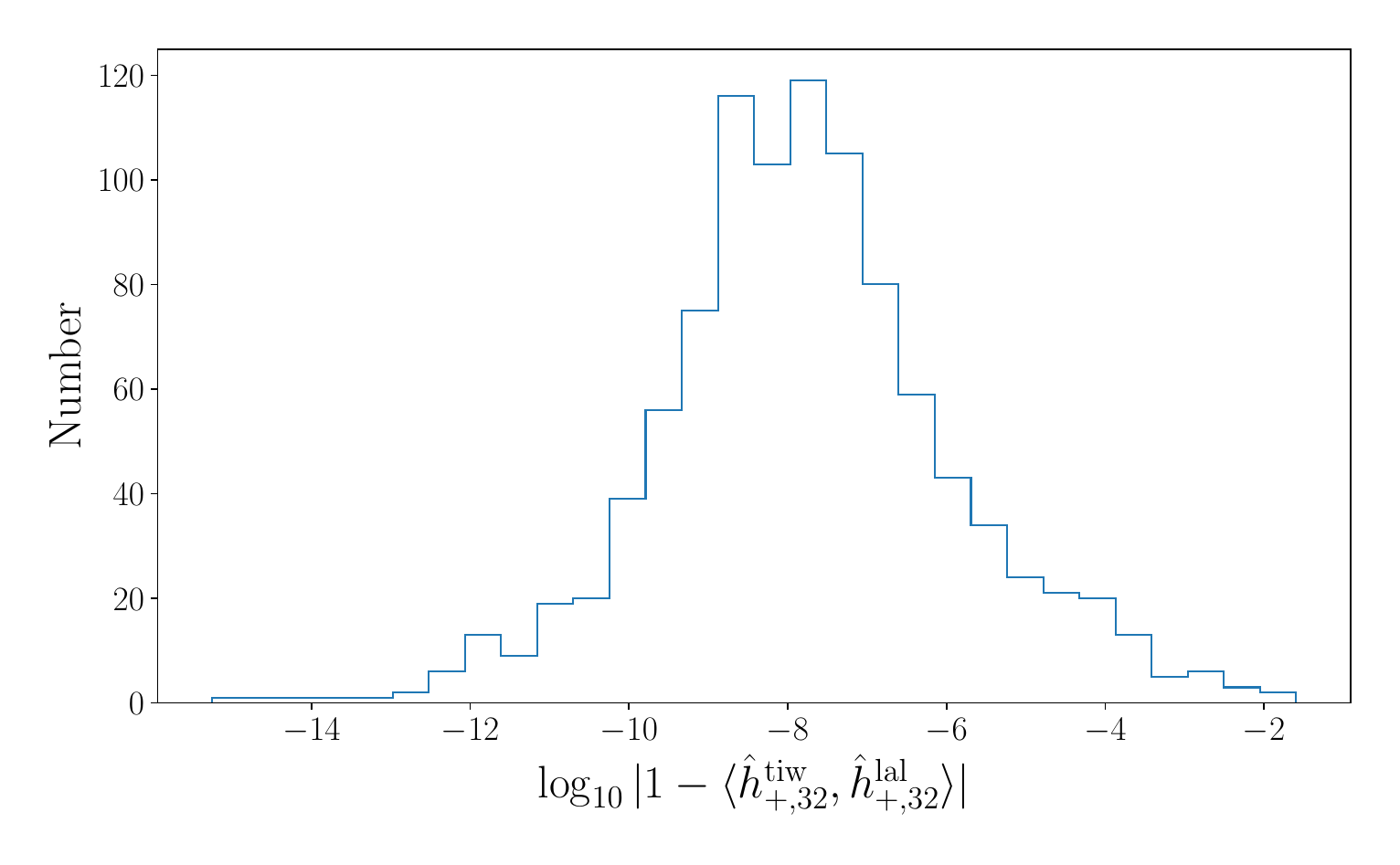}
    \caption{\textbf{Mismatch of the 32 mode in \texttt{PhenomXHM} between \texttt{tiwave} and \texttt{lalsimulation}.}
    For simplicity, we only consider the plus polarization here.
    We generate 1000 samples over the parameter space with $q\in[1,\ 1000],\ \chi_i\in[-0.99,\ 0.99]$ to compute the mismatch. Samples with exactly zero mismatch are excluded from the histograms, thus we use the number of samples rather than the density as the y-axis. In addition, due to numerical artifacts in the mismatch computation, a fraction of samples exhibit negative mismatch values. We therefore take the absolute value of the mismatch, which differs slightly from the commonly used definition given in Eq.~\ref{eq_mismatch}.
    }
    \label{fig_mism_32}
\end{figure}

We find the difference between waveforms generated by \texttt{tiwave} and \texttt{lalsimulation} are approach the limit of machine precision for 22, 21, 33, and 44 modes, where mismatches of all samples randomly selected in the parameter space are below $\mathcal{O}(10^{-15})$. 
However, waveforms of the 32 mode have relatively large differences, especially for samples of large mass ratio and high spins.
We demonstrate the mismatch of 32 mode in Fig.~\ref{fig_mism_32}. A total 1000 samples are generated over the parameter space of $q\in[1,\ 1000],\ \chi_i\in[-0.99,\ 0.99]$.
The fraction of samples where their mismatches are exactly zero is not counted in the histogram, thus we show the number rather the density in Fig.~\ref{fig_mism_32}.
Negative values of the mismatch may arise from numerical artifacts, the absolute value is therefore used. 
In addition, we only compare the plus polarization here for simplicity.
The differences mainly arise from the transition between the intermediate and merge-ringdown phase. 
To achieve a smooth transition, the first and second derivatives of the merge-ringdown phase are required.
However, these derivatives cannot be obtained analytically as in other modes, since the ansatz of merge-ringdown phase for 32 mode is constructed under the spheroidal-harmonic basis to better address the mode-mixing \citep{GarciaQuiros2020}, and the waveform under the spherical-harmonic basis is obtained by a linear transformation (more details can be found in Appendix A in reference \citep{GarciaQuiros2020}).
Thus, we can only get derivatives of merge-ringdown phase numerically by the finite difference scheme.
Catastrophic cancellation may occur when subtracting two nearby values, and can induce relatively large numerical errors.

\begin{figure}
    \centering
    \includegraphics[width=0.8\columnwidth]{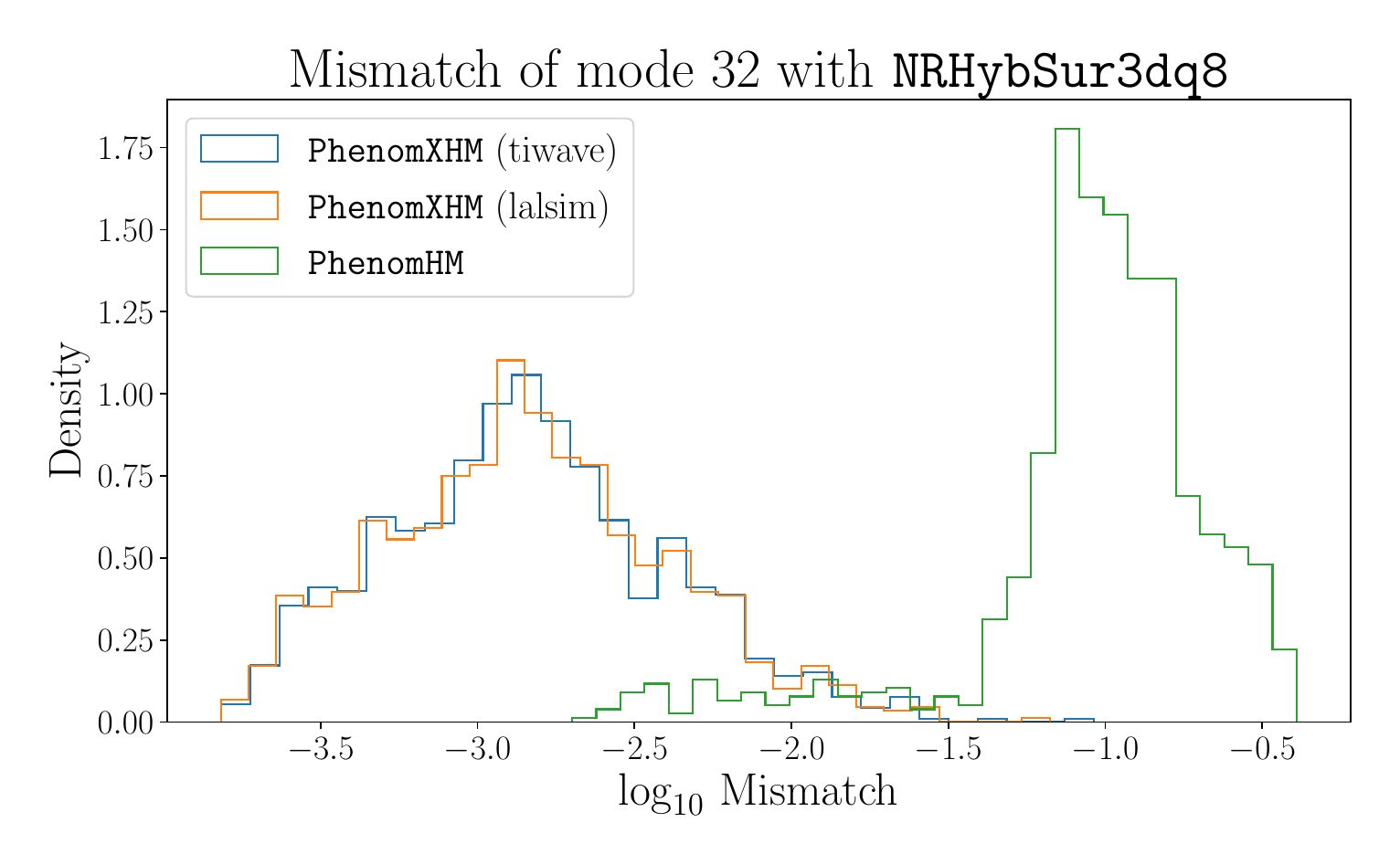}
    \caption{\textbf{Mismatch of the 32 mode against \texttt{NRHybSur3dq8} for \texttt{PhenomXHM} in \texttt{tiwave}, \texttt{PhenomXHM} in \texttt{lalsimulation}, and \texttt{PhenomHM}. }
    Samples are randomly generated over the parameter space allowed by \texttt{NRHybSur3dq8}, $q\in[1, 8],\ \chi_i\in[-0.8, 0.8]$. 
    Within this parameter space, the mismatch of the mode 32 for the reimplemented \texttt{PhenomXHM} in \texttt{tiwave} and for the waveform from \texttt{lalsimulation} is comparable.
    We believe that the reimplemented waveform in \texttt{tiwave} is reliable at least for moderate mass ratio and spins.
    }
    \label{fig_mism_32_nrhyb}
\end{figure}

However, we also compare the mismatch against the waveform \texttt{NRHybSur3dq8} over the parameter space of $q\in[1,\ 8],\ \chi_i\in[-0.8,\ 0.8]$ as shown in Fig.~\ref{fig_mism_32_nrhyb}, where mismatch of mode 32 in \texttt{IMRPhenomHM} is also shown as a reference.
It shows that in the parameter space allowed by \texttt{NRHybSur3dq8}, mismatches against \texttt{NRHybSur3dq8} of mode 32 are comparable for waveforms generated by \texttt{tiwave} and \texttt{lalsimulation}.
Thus, we believe \texttt{tiwave} is still safe to use for the situation of moderate mass ratio and spins.

\section{A minor modification in \texttt{PhenomXAS}} \label{app_mod_xas}
\begin{figure}
    \centering
    \includegraphics[width=0.8\columnwidth]{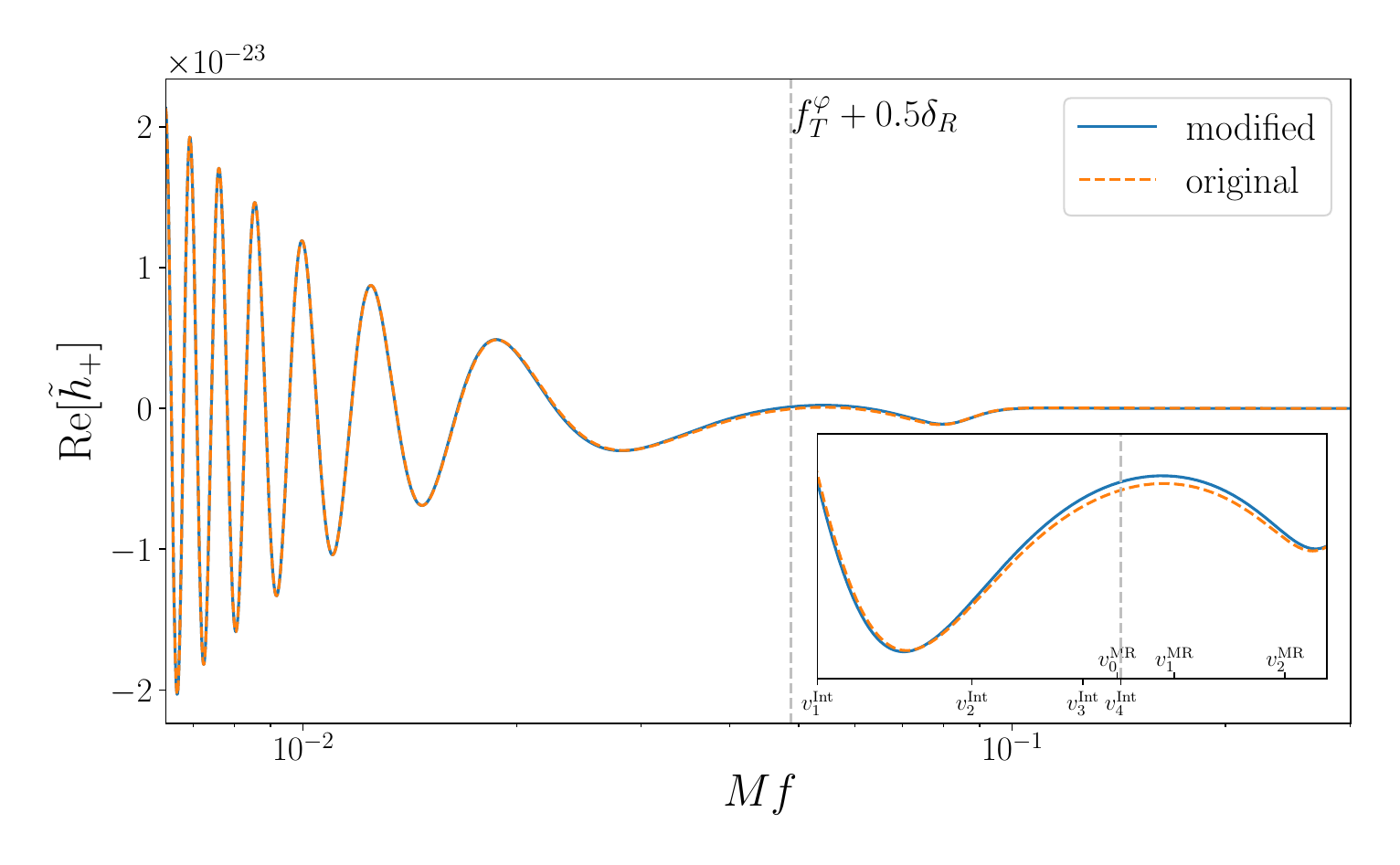}
    \caption{\textbf{Difference induced by the minor modification for an example signal.} 
    The blue solid line and the orange dashed line denote the modified and original waveforms of \texttt{PhenomXAS}, respectively. $v_i^{\mathrm{Int}}$ and $v_i^{\mathrm{MR}}$ in the inset denote the collocation points of intermediate phase and merge-ringdown phase.}
    \label{fig_mod_xas}
\end{figure}

\begin{figure}
    \centering
    \includegraphics[width=0.8\columnwidth]{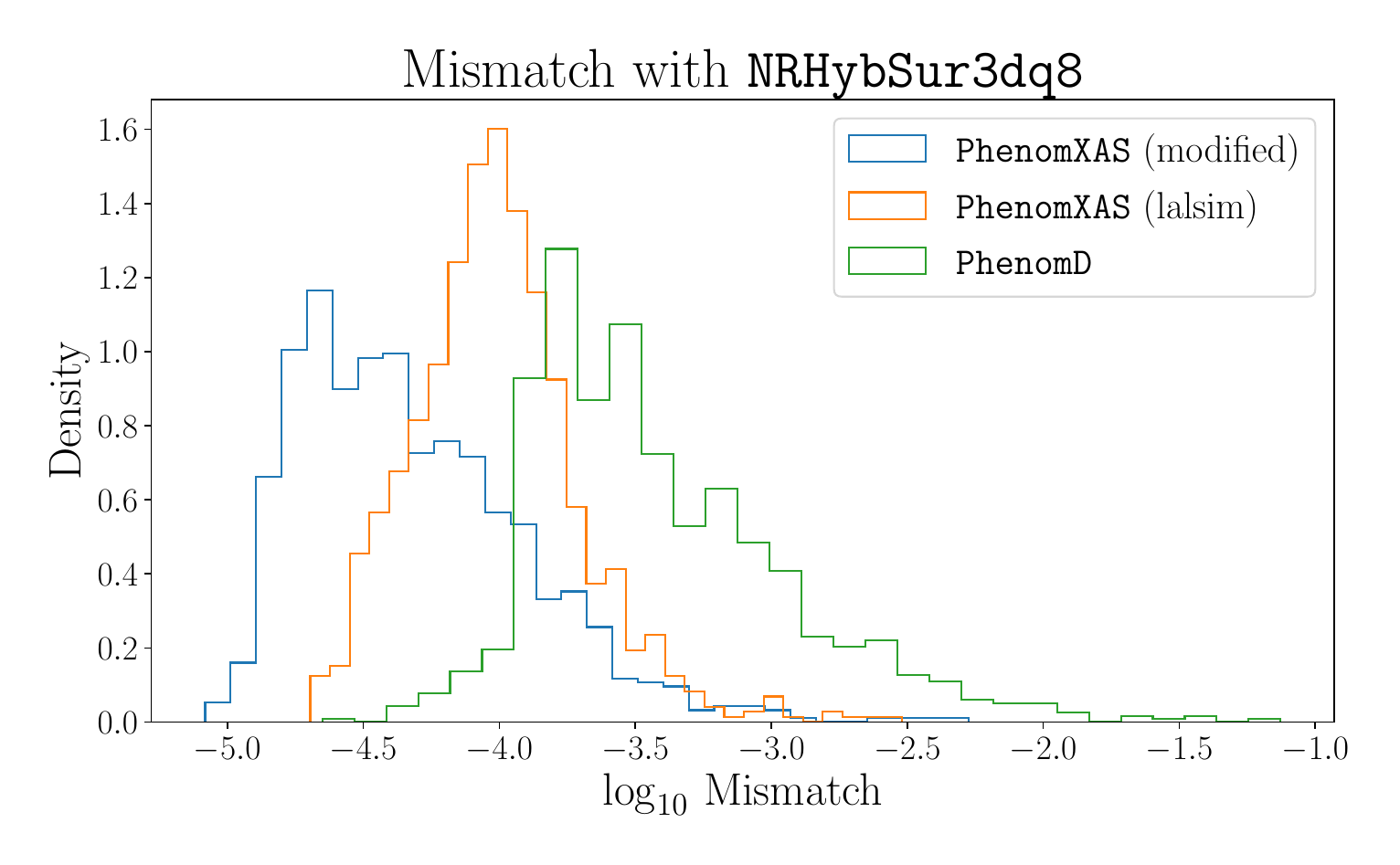}
    \caption{\textbf{Mismatch of \texttt{PhenomXAS} with the minor modification, original \texttt{PhenomXAS} in \texttt{lalsimulation}, and \texttt{PhenomD}, against the 22 mode in \texttt{NRHybSur3dq8}. }
    The histograms in blue, orange, and green colors denote mismatch of modified \texttt{PhenomXAS}, original \texttt{PhenomXAS}, and \texttt{PhenomD}, respectively. Samples are generated in the parameter space of $q\in[1, 8],\ \chi_i\in[-0.8, 0.8]$. Waveforms generated by \texttt{lalsimulation} use the default setting with multibanding disabled.
    }
    \label{fig_mism_xas}
\end{figure}

During the reimplementation of the waveform \texttt{PhenomXAS}, we find that a minor modification can slightly improve its mismatch against the 22 mode in \texttt{NRHybSur3dq8}.
In the construction of intermediate phase, a system of linear equations is solved to determine the values of the ansatz parameters, where the coefficient matrix is determined by the collocation points and the form of the ansatz, the right-hand-side vector is set by fitted values at the collocation points.
In \texttt{lalsimulaiton}, the value at the last collocation point of intermediate phase is set by the value at the first collocation point of merge-ringdown phase (see \href{https://git.ligo.org/lscsoft/lalsuite/-/blob/e59d1f0003d2358dd49b9795073624fdcf1ce0a8/lalsimulation/lib/LALSimIMRPhenomX_internals.c#L2144}{L. 2144} and \href{https://git.ligo.org/lscsoft/lalsuite/-/blob/e59d1f0003d2358dd49b9795073624fdcf1ce0a8/lalsimulation/lib/LALSimIMRPhenomX_internals.c#L1205}{L. 1205} in \texttt{LALSimIMRPhenomX\_internals.c}).
However, as discussed in the reference \citep{Pratten2020}, these two collocation points are not identical. The last collocation point of intermediate phase is taken to be $f^\varphi_T+0.5\delta_R$, whereas the first collocation point of merge-ringdown phase is taken to be $f^\varphi_T$ (the definition of $f^\varphi_T$ and $\delta_R$ can be found in Eq. 5.10 and 5.11 of \citep{Pratten2020}).
Therefore, we make a minor modification where the value at last collocation point for intermediate phase is set by the value of merge-ringdown phase computed at $f^\varphi_T+0.5\delta_R$.
We use an example signal to show the difference induced by this minor modification in Fig.~\ref{fig_mod_xas}.
The modification can slightly improve the mismatch against \texttt{NRHybSur3dq8} as shown in Fig.~\ref{fig_mism_xas}.
Note that this minor modification is considered only in this appendix, in all other part of this paper, the waveforms of \texttt{PhenomXAS} or the 22 mode in \texttt{PhenomXHM} are generated strictly following the implementation in \texttt{lalsimulation}.

\begin{acknowledgments}
This work is supported by the National Key R\&D Program of China (Grant No.~2022YFC2204603 and 2022YFC2204602), the National Natural Science Foundation of China (Grant No.~12325301, 12273035, and 12405075), Guizhou Provincial Major Scientific and Technological Program XKBF (2025)011.
The numerical calculations in this paper have been done on the supercomputing system in the Supercomputing Center of University of Science and Technology of China.
Data analyses and results visualization in this work made use of \texttt{bilby} \citep{Ashton2019}, \texttt{multinest} \citep{Feroz2007,Feroz2008,Feroz2013}, \texttt{lalsuite} \citep{lalsuite}, \texttt{numpy} \citep{Harris2020, Walt2011}, \texttt{scipy} \citep{Virtanen2020}, and \texttt{matplotlib} \citep{Hunter2007}.
\end{acknowledgments}

\bibliography{ref}
\bibliographystyle{aasjournalv7}

\end{document}